\documentclass[a4paper,aps,twocolumn,pre,intlimits]{revtex4-1}
\usepackage{amsmath,amsthm,amssymb,amsthm,dsfont,mathrsfs,bbm}
\usepackage{mathtools}
\usepackage{multirow}
\usepackage[usenames,dvipsnames]{xcolor}
\usepackage{bbm,bm,mathrsfs,dsfont}
\usepackage[english]{babel}
\usepackage{pgfplotstable}
\usepackage{hyperref}
\hypersetup{pdfauthor={Caracciolo DAchille Malatesta Sicuro},pdftitle={Finite size corrections},%
            colorlinks, linktocpage=true, pdfstartpage=3, pdfstartview=FitV,%
    urlcolor=orange, linkcolor=blue, citecolor=red, 
        }
\usepackage[T1]{fontenc}
\usepackage{newtxtext,newtxmath}

\newcommand{\mG}{\mathcal{G}}

\newcommand{\R}{\mathds{R}}

\DeclareMathOperator{\dd}{d}
\DeclareMathOperator{\DD}{\mathcal D}
\newcommand{\fder}[2]{\DD_{#2}^{#1}}
\DeclareMathOperator{\e}{e}

\allowdisplaybreaks
\begin{document}
\title{Finite size corrections in the random assignment problem}
\author{Sergio Caracciolo}\email{sergio.caracciolo@mi.infn.it}
\affiliation{Dipartimento di Fisica, University of Milan and INFN, via Celoria 16, 20133 Milan, Italy}
\author{Matteo P. D'Achille}\email{matteopietro.dachille@studenti.unimi.it}
\affiliation{Dipartimento di Fisica, University of Milan and INFN, via Celoria 16, 20133 Milan, Italy}
\author{Enrico M. Malatesta}\email{enrico.m.malatesta@gmail.com}
\affiliation{Dipartimento di Fisica, University of Milan and INFN, via Celoria 16, 20133 Milan, Italy}
\author{Gabriele Sicuro}\email{gabriele.sicuro@roma1.infn.it}
\affiliation{Dipartimento di Fisica, Sapienza Universit\`a di Roma, P.le A. Moro 2, 00185, Rome, Italy}

\date{\today}
\begin{abstract}We analytically derive, in the context of the replica formalism, the first finite size corrections to the average optimal cost in  the random assignment problem for a quite generic distribution law for the costs. We show that, when moving from a power-law distribution to a $\Gamma$ distribution, the leading correction changes both in sign and in its scaling properties. We also examine the behavior of the corrections when approaching a $\delta$-function distribution. By using a numerical solution of the saddle-point equations, we provide predictions that are confirmed by numerical simulations.
\end{abstract}
\maketitle

\section{Introduction}\label{sec:intro}
Matching is a classical problem in combinatorial optimization \cite{Papadimitriou1998a,lovasz2009matching}. It can be defined on any \textit{graph} $\mathcal G=(\mathcal V, \mathcal E)$, where $\mathcal V$ is its set of \textit{vertices} and $\mathcal E$ its set of \textit{edges}. A \textit{matching} on $\mathcal G$ is a set of non adjacent edges of $\mathcal G$, that is, a set of edges that do not share a common vertex. A matching is \textit{maximal} when the addition of any new edge to the set makes it no longer a matching. A matching is said to be \textit{maximum} when it has the maximal cardinality among the maximal matchings. All the maximum matchings have the same cardinality $\nu(\mathcal G)$, which is called the \textit{matching number} of the graph $\mathcal G$.
A {\em perfect matching} (or {\em 1-factor}) is a matching that matches all vertices of the graph. That is, every vertex of the graph is incident to exactly one edge of the matching. Every perfect matching is maximum and hence maximal. A perfect matching is also a minimum-size edge cover. We will denote by $\mathcal M$ the set of perfect matchings.

Suppose now that we can assign a \textit{cost} $w_e \ge 0$ to each edge $e\in \mathcal E$. For each perfect matching $\pi\in\mathcal M$ we define a total cost (energy)
\begin{equation}
E(\pi)\coloneqq\sum_{e \in \pi} w_e
\end{equation}
and a mean cost for edge
\begin{equation}
\epsilon(\pi)\coloneqq \frac{1}{\nu(\mathcal G)} \sum_{e \in \pi} w_e.
\end{equation}
The \textit{matching problem} consists in the choice of the optimal perfect matching $\pi^*\in\mathcal M$ which minimizes the total cost
\begin{equation}
E(\pi^*) = \min_{\pi \in\mathcal M} E(\pi).
\end{equation}
Let us now consider the matching problem on the complete bipartite graph $\mathcal {K}_{N,M}$, in which the vertex set is the union of two disjoint sets $\mathcal V_1$ and $\mathcal V_2$ of cardinality $N$ and $M$, respectively. Defining $[n]\coloneqq\{1,\dots,n\}$, we identify $\mathcal V_1$ and $\mathcal V_2$ with $[N]$ and $[M]$, respectively, and the edge set is therefore the set of all couples $e=(i,j)$ with $i\in[N]$ and $j\in [M]$. In this case the matching problem is usually called the (general) \textit{assignment} problem and the matching number is $\nu(\mathcal{K}_{N,M})=\min \{N, M\}$. In the following we will concentrate on the $N=M$ case, in which a perfect matching $\pi$ is a permutation in the symmetric group $\mathcal{S}_N$ and can be represented by a square matrix with entries $\pi_{ij}\in\{0, 1\}$ for all $i\in [N]$ and $j\in [N]$ such that
\begin{equation}
\pi_{ij} = 
\begin{cases}
1 & \text{for } e=(i,j) \in \pi \\
0 & \text{otherwise},
\end{cases} \label{pi}
\end{equation}
with the constraints
\begin{equation} \label{constraints}
\sum_{i=1}^{N} \pi_{ij} = \sum_{i=1}^{N} \pi_{ji} = 1\quad \forall j\in[N].
\end{equation}
The matching cost associated with $\pi$ can be written as
\begin{equation}\label{E}
E(\pi) = \sum_{i=1}^N  \sum_{j=1}^N \pi_{ij} w_{ij}.
\end{equation}
From the point of view of computational complexity, matching problems are \textit{simple} problems, being in the \texttt{P} complexity class, as \textcite{Kuhn} proved with his celebrated Hungarian algorithm for the assignment problem. Very fast algorithms are nowadays available both to find perfect matchings and to solve the matching problem on a generic graph \cite{Edmonds1965,edmonds1972,Micali1980,lovasz2009matching}. 

The properties of the solution of a matching problem on a given ensemble of realizations are often interesting as well \cite{Steele1997}. In the {\em random assignment problem}, for example, we consider the matching problem on $\mathcal {K}_{N,N}$, whereas the costs for all edges are independent random variables, identically distributed with a common law $\rho$. Each particular choice $\mathcal W=\{w_e\}_{e\in\mathcal E}$ for the set of edge costs is called an \textit{instance} of the problem. In this random version of the problem, we are interested in the \textit{typical} properties of the optimal matching. In particular we will concentrate on the asymptotic behavior for large $N$ of the average optimal cost
\begin{equation}
\overline{E}\coloneqq\overline{ E(\pi^*)}=\overline{\min_{\pi\in\mathcal M}\sum_{i=1}^N  \sum_{j=1}^N \pi_{ij} w_{ij}},
\end{equation}
where we have denoted by an over bar the average over all possible instances (i.e., the average over the \textit{disorder}). In the same way, we can consider the matching problem with random weights on the complete graph $\mathcal K_{2N}$, having $2N$ vertices such that each one of them is connected to all the others. We simply call this variation of the problem \textit{random matching problem}. Both the random matching problem and the random assignment problem have been solved by \textcite{Mezard1985} by means of the replica trick. The random assignment problem and the random matching problem have also been generalized to the Euclidean case, in which the weights in $\mathcal W$ are functions of the distances between points associated with the vertices of the graph and the points are assumed to be randomly generated on a certain Euclidean domain \cite{Shor1985,Mezard1988,Yukich1998,Lucibello2017}. Due to the underlying Euclidean structure, dimensionality plays an important role in the scaling of the optimal cost of random Euclidean matching problems \cite{Mezard1988,Caracciolo2014}, and correlation functions can be introduced and calculated \cite{Boniolo2012,Caracciolo2015a}. Euclidean matching problems proved to be deeply connected with Gaussian stochastic processes \cite{Boniolo2012,Caracciolo2014c} and with the theory of optimal transport \cite{Caracciolo2015}. In the latter context, \textcite{Ambrosio2016} rigorously derived the asymptotic behavior of the average optimal cost for the two-dimensional random Euclidean assignment problem, previously obtained in Ref.~\cite{Caracciolo2014} using a proper scaling ansatz. For a recent review on random Euclidean matching problems, see Ref.~\cite{Sicuro2017}.

Remarkably enough, after the seminal works of \textcite{Kirkpatrick1983}, \textcite{Orland1985}, and M\'ezard and Parisi, the application of statistical physics techniques to random optimization problems proved to be extremely successful in the study of the {typical} properties of the solutions, but also in the development of algorithms to solve a given instance of the problem \cite{mezard2009information,Bapst2013}. In formulating a combinatorial problem as a model in statistical mechanics, an artificial inverse temperature $\beta$ is introduced to define a Boltzmann weight $\exp\left(-\beta E\right)$ for each configuration. Of course, configurations of minimal energy are the only ones to contribute in the limit of infinite $\beta$. For example, in the assignment problem, the corresponding partition function for each instance is
\begin{equation}
Z[w] 
=\sum_\pi  \left[ \prod_{j=1}^N \delta \left( 1- \sum_{i=1}^N \pi_{ij}\right) \delta \left( 1- \sum_{i=1}^N \pi_{ji}\right) \right] \, \e^{- \beta E(\pi)},
\end{equation}
where the ``energy'' $E(\pi)$ is given by~\eqref{E}. Thermodynamic information is obtained from the average total free energy
\begin{align}
\overline{F}&\coloneqq - \frac{\overline{\ln Z}}{\beta},\\
\overline{E}&=\frac{\partial}{\partial \beta}  \beta\overline{F}.
\end{align}
In this paper we apply the formalism above to the {random} assignment problem, where the costs of all the edges are taken to be independent and identically distributed random variables with probability distribution density $\rho_r(w)$ such that, in the neighborhood of the origin, $\rho_r$ can be written as
\begin{equation}
\rho_r(w)=w^r\sum_{k=0}^\infty \eta_k(r) w^k,\quad r>-1,\quad \eta_0(r)\neq 0.\label{pr}
\end{equation}
In the previous expression, $\eta_k(r)$ are coefficients (possibly dependent on $r$) of the Maclaurin series expansion of the function $\rho_r(w)w^{-r}$, which is supposed to be analytic in the neighborhood of the origin. The constraint $r>-1$ is required to guarantee the integrability of the distribution near the origin. By the general analysis performed in Refs.~\cite{Orland1985,Mezard1985}, which we will resume in Section~\ref{sec:SR}, the average cost, in the asymptotic regime of an infinite number $N$ of couples of matched points, will depend on the power $r$ that appears in Eq.~\eqref{pr} only, aside from a trivial overall rescaling related to $\eta_0$. More precisely, if $\overline{E}_r$ is the average optimal cost obtained using the law $\rho_r$, then
\begin{subequations}
\begin{equation}
\hat{E}_r = \lim_{N\to \infty} \frac{1}{N^\frac{r}{r+1}}\overline{E}_r = \frac{r+1}{\left[\eta_0 \Gamma(r+1)\right]^\frac{1}{r+1}}J_r^{(r+1)} \label{Er}
\end{equation}
where 
\begin{equation}\label{Jint}
J_r^{(\alpha)} \coloneqq \int_{-\infty}^{+\infty} \hat G_r(-u) \fder{\alpha}{u}\hat G_r(u)\dd u
\end{equation}
(we will later specify the meaning of the fractional order derivative $\fder{\alpha}{u}$). The function $\hat G_r(y)$ is the solution of the integral equation
\begin{equation}
\hat{G}_r(l) =  \int_{-l}^{+\infty} \frac{(l+y)^r}{\Gamma(r+1)}\e^{-\hat{G}_r(y)} \, dy \label{Gr}
\end{equation}
\end{subequations}
and it is analytically known for $r=0$ and, after a proper rescaling of its variable, in the $r\to \infty$ limit.

Our main results concern the finite-size corrections to the average optimal costs, and they will be presented in Section~\ref{sec:fsc}, extending the classical achievements in Refs.~\cite{Mezard1987,Ratieville2002}. In particular, we obtain the expansion\begin{subequations} \label{resume}
\begin{equation}
\hat{E}_r(N)=\hat{E}_r +\sum_{k=1}^{[r]+1}\Delta\hat{F}_r^{(k)} +\Delta\hat{F}_r^T + \Delta\hat{F}_r^F+o\left(\frac{1}{N}\right),
\end{equation}
where $[r]$ is the integer part of $r$, $[r]\leq r<[r]+1$ (the sum is absent for $r<0$), and the corrections have the structure
\begin{align}
\Delta\hat{F}_r^{(k)} = &\frac{\Delta\phi^{(k)}_r}{N^\frac{k}{r+1}},\quad r\geq 0,\quad 1\leq k\leq [r]+1,\\
\Delta\hat{F}_r^T = &- \frac{1}{N} \, \frac{\Gamma(2r+2)J_r^{(0)}}{(r+1) \eta_0^\frac{1}{r+1} [\Gamma(r+1)]^\frac{2 r+3}{r+1}}\\
\Delta\hat{F}_r^F = & - \frac{1}{N} \frac{1}{2\, [\eta_0 \Gamma(r+1)]^\frac{1}{r+1}}\frac{1}{J_r^{(r+3)}},
\end{align}
$\Delta\phi^{(k)}_r$ being independent of $N$. In particular, for $r> 0$, we have that, provided $\eta_1\neq 0$, the first finite-size correction is given by
\begin{equation}
\Delta\hat{F}_r^{(1)} = - \frac{\eta_1}{N^\frac{1}{r+1}}\frac{r+1}{\eta_0[\eta_0 \Gamma(r+1)]^\frac{2}{r+1}}J_r^{(r)}.\label{DF1}
\end{equation}
\end{subequations}

In our discussion, we will consider in particular two probability distribution densities, namely, the Gamma distribution
\begin{equation}\rho_r^\Gamma(w)\coloneqq \frac{w^r \e^{-w} \theta(w)}{\Gamma(r+1)},\end{equation}
defined on $\R^+$, and the power-law distribution
\begin{equation}
\rho_r^\text{P}(w)\coloneqq (r+1)w^r\theta(w)\theta(1-w),\end{equation}
defined on the compact interval $[0,1]$. In the previous expressions we have denoted by $\theta(w)$ the Heaviside theta function on the real line. Observe that, for the distribution $\rho_r^\Gamma$, we have
\begin{equation}
\eta^\Gamma_k(r)=\frac{1}{\Gamma(r+1)}\frac{(-1)^k}{k!},\quad k\geq 0,\label{cGamma}
\end{equation}
whereas in the case of $\rho_r^\text{P}$,
\begin{equation}
\eta_k^\text{P}(r)=(r+1)\, \delta_{k,0},\quad k\geq 0.\label{cP}
\end{equation}
The case $r=0$ has already been considered by \textcite{Mezard1987} and subsequently revised and corrected by \textcite{Ratieville2002}. In the case analyzed in their works, the contributions $\Delta\hat{F}_0^{(1)}$, $\Delta\hat{F}_0^T$, and $\Delta\hat{F}_0^F$ are of the same order. This is not true anymore for a generic distribution with $r\neq 0$. As anticipated, a relevant consequence of our evaluation is that, if $\eta_1\neq 0$, for $r>0$ the most important correction comes from $\Delta\hat{F}_r^{(1)}$ and scales as $N^{-\frac{1}{r+1}}$. It follows that, in order to extrapolate to the limit of an infinite number of points, the best choice for the law for random links (in the sense of the one that provides results closer to the asymptotic regime) is the pure power law $\rho_r^\text{P}$, where only analytic corrections in inverse power of $N$ are present. Such a remark is even more pertinent in the limit when $r\to \infty$ at a fixed number of points, where the corrections $\Delta\hat{F}_r^{(k)}$ become of the same order of the leading term. Indeed, the two limits $r\to \infty$ and $N\to \infty$ commute only if the law $\rho_r^\text{P}$ is considered.

The paper is organized as follows. In Section~\ref{sec:SR} we review, in full generality, the calculation of the replicated partition function of the random assignment problem. In Section~\ref{sec:fsc} we evaluate the finite-size corrections, discussing the different contributions and proving Eqs.~\eqref{resume}. In Section~\ref{sec:rinf} we evaluate the relevant $r\to\infty$ case, pointing out the noncommutativity of this limit with the thermodynamic limit. In Section~\ref{sec:num} we provide the numerical values of the necessary integrals and we compare our prediction with a Monte Carlo simulation for different values of $r$. In Section~\ref{sec:conc} we summarize and give our conclusions.

\section{The replicated action}\label{sec:SR}
In the present section we perform a survey of the classical replica computation for the random assignment problem, following the seminal works of \textcite{Mezard1985,Mezard1987} (for a slightly different approach see also Ref.~\cite{Orland1985}), but we do not adopt their choice to replace $\beta$ with $\beta/2$. As anticipated in Section \ref{sec:intro}, the computation of the average of $\ln Z$ goes through the replica trick \cite{Edwards1975}
\begin{equation}
\overline{\ln Z} = \lim_{n \to 0} \frac{\overline{Z^n} -1}{n}.
\end{equation}
In other words, in order to compute $\overline{\ln Z}$ we introduce $n$ noninteracting replicas of the initial system, denoted by the index $a\in[n]$. For each $i\in[N]$, $2n$ replicated fields $\{\lambda_i^a,\mu_i^a\}_{a=1,\dots,n}$ appear to impose the constraints in Eq.~\eqref{constraints}, using the relation
\begin{equation}
 \int_0^{2\pi}\e^{ik\lambda}\dd\lambda=2\pi\delta_{k\,0}.\label{lagrangemultiplier}
\end{equation}
We obtain
\begin{multline}
Z^n[w] = \left[\prod_{a=1}^n \prod_{i=1}^N  \int_0^{2 \pi} \frac{\dd \lambda_i^a} {2 \pi} \int_0^{2 \pi} \frac{\dd \mu_i^a} {2 \pi} \e^{i (\lambda_i^a + \mu_i^a)} \right]\times\\\times
\prod_{i=1}^N  \prod_{j=1}^N \prod_{a=1}^n\left[ 1 + \e^{- i ( \lambda_j^a + \mu_i^a) - \beta w_{ij}  } \right].
\end{multline}
Let ${\cal P}([n])$ be the set of subsets of the set $[n]$ and for each subset $\alpha\in {\cal P}([n])$ let $|\alpha|$ be its cardinality. Then
\begin{multline}
\prod_{a=1}^n\left[ 1 + \e^{- i ( \lambda_j^a + \mu_i^a) - \beta w_{ij}  } \right] \\
=\sum_{\alpha\in {\cal P}([n])} \e^{- \beta |\alpha| w_{ij}- i \sum_{a\in \alpha} ( \lambda_j^a + \mu_i^a )}\\
= 1 + \sum_{p=1}^n \e^{-\beta\, p \, w_{ij}}\sum_{\substack{\alpha\in \mathcal P([n])\\|\alpha|=p}}\e^{-i \sum_{a\in\alpha} ( \lambda_j^a + \mu_i^a )},
\end{multline}
where we have extracted the contribution from the empty set in the sum, which is $1$, and we have partitioned the contribution from each subset of replicas in terms of their cardinality. This expression is suitable for the average on the costs. From the law $\rho_r$ we want to extract the leading term for large $\beta$ of the contribution of each subset $\alpha\in {\cal P}([n])$ with $|\alpha|=p$. In particular, we define
\begin{equation}
g_{\alpha}\equiv g_{p}\coloneqq \int_0^{+\infty} \rho_r(w) \e^{- \beta p w}\dd w.\label{ga}
\end{equation}
Due to the fact that short links only participate in the optimal configuration, approximating $\rho_r(w)\sim\eta_0 w^r$, we obtain that the minimal cost for each matched vertex is of the order $N^{-\frac{1}{r+1}}$ so the total energy $E$ and the free energy should scale as $N^{\frac{r}{r+1}}$, that is, the limits
\begin{align}
\lim_{N\to \infty} \frac{1}{N^\frac{r}{r+1}}\overline{F} = &  \hat{F}, \\
\lim_{N\to \infty} \frac{1}{N^\frac{r}{r+1}}\overline{E} = &  \hat{E} \, 
\end{align}
are finite. This regime can be obtained by considering in the thermodynamic limit
\begin{equation}
\beta =\hat\beta N^\frac{1}{r+1}   \label{beta},
\end{equation}
where $\hat{\beta}$ is kept fixed. As a consequence we set
\begin{multline}\label{ghat}
\hat{g}_p\coloneqq N g_p=N\int_0^{+\infty} \rho_r(w) \e^{- p\hat\beta N^\frac{1}{r+1} w}\dd w\\
=\sum_{k=0}^{+\infty}\frac{1}{N^{\frac{k}{r+1}}}\frac{\eta_k\Gamma(k+r+1)}{(\hat \beta p)^{k+r+1}}.
\end{multline}
The replicated partition function can be written therefore as
\begin{multline}\label{Zn}
\overline{Z^n}=\left[\prod_{a=1}^n \prod_{i=1}^N  \int_0^{2 \pi} \frac{\dd \lambda_i^a} {2 \pi} \int_0^{2 \pi} \frac{\dd \mu_i^a} {2 \pi}\,  \e^{i (\lambda_i^a + \mu_i^a)} \right] \,
\prod_{i,j}^N \left( 1 + \frac{T_{ij}}{N} \right)\\
=\left[\prod_{a=1}^n \prod_{i=1}^N  \int_0^{2 \pi} \frac{\dd \lambda_i^a} {2 \pi} \int_0^{2 \pi} \frac{\dd \mu_i^a} {2 \pi}\,  \e^{i (\lambda_i^a + \mu_i^a)} \right]\times\\\times \exp\left[\frac{1}{N} 
\sum_{i=1}^N\sum_{j=1}^N \left(T_{ij}-\frac{T_{ij}^2}{2N}\right)+o\left(\frac{1}{N^2}\right)\right],
\end{multline}
with
\begin{equation}
T_{ij}\coloneqq \sideset{}{'}\sum_{\alpha\in {\cal P}([n])}\hat{g}_{\alpha}\e^{- i \sum_{a\in \alpha} ( \lambda_j^a + \mu_i^a )}
\end{equation}
where in the sum $\sum'$ on subsets the empty set is excluded. If we introduce, for each subset $\alpha\in {\cal P}([n])$, the quantities
\begin{subequations}
\begin{align}
\frac{x_\alpha + i \, y_\alpha}{\sqrt{2}} :=  & \sum_{k=1}^N \e^{- i \sum_{a\in \alpha} \lambda_k^a} \label{x1} \\
\frac{x_\alpha - i \, y_\alpha}{\sqrt{2}} :=  & \sum_{k=1}^N \e^{- i \sum_{a\in \alpha} \mu_k^a} \label{x2}
\end{align}\end{subequations}
we can write
\begin{subequations}\begin{align}
\sum_{i=1}^N  \sum_{j=1}^N  T_{ij}&= 
\sideset{}{'}\sum_{\mathclap{\alpha\in {\cal P}([n])}}\hat{g}_{\alpha}\,  \frac{x_\alpha^2 + y_\alpha^2}{2},\\
\sum_{i=1}^N  \sum_{j=1}^N  T^2_{ij}&= 
\sideset{}{'}\sum_{\mathclap{\alpha,\beta\in {\cal P}([n])}}\hat{g}_{\alpha}\hat{g}_{\beta} \frac{x_{\alpha\cup\beta}^2 + y_{\alpha\cup\beta}^2}{2}.\label{29b}
\end{align}
As observed by \textcite{Mezard1987} and \textcite{Ratieville2002}, in Eq.~\eqref{29b} we can constrain the sum on the right-hand side to the couples $\alpha,\beta\in\mathcal P([n])$ such that $\alpha\cap\beta=\emptyset$. Indeed, let us consider $\alpha,\beta\in\mathcal P([n])$ and $\alpha\cap\beta\neq\emptyset$. Then, defining $\alpha\triangle\beta	\coloneqq(\alpha\cup\beta)\setminus(\alpha\cap\beta)$, we have that
\begin{multline}
 \frac{x_{\alpha\cup\beta}^2+y_{\alpha\cup\beta}^2}{2}=\\
 =\sum_{l,k}\exp\left[-2i\sum_{\mathclap{a\in\alpha\cap\beta}}\left(\lambda_l^a+\mu_k^a\right)-i\sum_{\mathclap{b\in\alpha\triangle\beta}}\left(\lambda_l^b+\mu_k^b\right)\right].
\end{multline}
Due to Eq.~\eqref{lagrangemultiplier} and to the presence of the coefficients $\exp\left(-2i\lambda_l^a-2i\mu_k^a\right)$, the contribution of the term above will eventually be suppressed because of the integration over the Lagrange multipliers in the partition function. We can therefore simplify our calculation by substituting immediately 
\begin{equation}
 \sum_{i=1}^N  \sum_{j=1}^N  T^2_{ij}= 
\sideset{}{'}\sum_{\substack{\alpha,\beta\in {\cal P}([n])\\\alpha\cap\beta=\emptyset}}\hat{g}_{\alpha}\hat{g}_{\beta} \frac{x_{\alpha\cup\beta}^2 + y_{\alpha\cup\beta}^2}{2}.
\end{equation}
\end{subequations}
We perform now a Hubbard--Stratonovich transformation, neglecting $o(N^{-2})$ terms in the exponent in Eq.~\eqref{Zn}, obtaining
\begin{multline}
\exp\left[\frac{1}{N} 
\sum_{i=1}^N\sum_{j=1}^N \left(T_{ij}-\frac{T_{ij}^2}{2N}\right)\right]=\\
= \left[\sideset{}{'}\prod_{\alpha\in {\cal P}([n])}\iint\frac{N\dd X_\alpha \dd Y_\alpha}{2 \pi\hat{g}_{\alpha} }\exp\left(x_\alpha X_\alpha + y_\alpha Y_\alpha\right)\right]\times\\\times 
\exp\left[-N\sideset{}{'}
\sum_{\mathclap{\alpha\in {\cal P}([n])}}\frac{X_\alpha^2 + Y_\alpha^2}{2\hat{g}_{\alpha}}-\sideset{}{'}\sum_{\mathclap{\substack{\alpha,\beta\in {\cal P}([n])\\\alpha\cap\beta=\emptyset}}}\hat{g}_{\alpha}\hat{g}_{\beta} \frac{X_{\alpha\cup\beta}^2 + Y_{\alpha\cup\beta}^2}{4\hat g_{\alpha\cup\beta}^2}\right],
\end{multline}
up to higher order terms in the exponent. Now, let us observe that
\begin{multline}
x_\alpha X_\alpha + y_\alpha Y_\alpha =
\left( \sum_{i=1}^N \e^{- i \sum_{a\in \alpha} \lambda_i^a}  \right) \frac{X_\alpha - i \, Y_\alpha}{\sqrt{2}}\\
+\left( \sum_{i=1}^N \e^{- i \sum_{a\in \alpha} \mu_i^a}  \right) \frac{X_\alpha +i \, Y_\alpha}{\sqrt{2}}.
\end{multline}
Introducing the function of $v_\alpha$,
\begin{equation}
z[v_\alpha]\coloneqq\left[\prod_{a=1}^n\int_0^{2 \pi} \frac{\dd \lambda^a}{2 \pi}\e^{i \lambda^a} \right]\exp \left[v_\alpha \e^{- i \sum_{b\in \alpha} \lambda^b}   \right],
\end{equation}
and the order parameters
\begin{equation}
Q_\alpha\coloneqq\frac{X_\alpha +iY_\alpha}{\sqrt{2}},
\end{equation}
we can write
\begin{subequations}\begin{multline}\label{Zn2}
\overline{Z^n} =\\= \left[\sideset{}{'} \prod_{\alpha\in {\cal P}([n])}
 \frac{N}{ 2 \pi\, \hat{g}_{\alpha} }\iint \dd Q_\alpha \dd Q^*_\alpha\right]\e^{-NS[Q]-N\Delta S^T[Q]},
\end{multline}
with
\begin{align}
S[Q]&= \sideset{}{'}\sum_{\mathclap{\alpha\in {\cal P}([n])}}\;\; \left(\frac{|Q_\alpha|^2}{\hat{g}_{\alpha} } - \ln z\left[ Q_\alpha \right]  - \ln z\left[Q^*_\alpha\right]\right),
\label{S}\\
\Delta S^T[Q]&=\sideset{}{'}\sum_{\mathclap{\substack{\alpha,\beta\in {\cal P}([n])\\\alpha\cap\beta=\emptyset}}}\hat g_\alpha\hat g_\beta\frac{|Q_{\alpha\cup\beta}|^2}{ 2N\hat{g}^2_{\alpha\cup\beta}},\label{DST}
\end{align}\end{subequations}
a form that is suitable to be evaluated, in the asymptotic limit for large $N$, by means of the saddle-point method. It is immediately clear that $\Delta S^T$ contains a contribution to the action that is $O(N^{-1})$ and therefore it can be neglected in the evaluation of the leading contribution. It follows that the stationarity equations are of the form
\begin{equation}
\frac{Q^*_\alpha}{\hat{g}_{\alpha} } =\frac {\dd \ln z[Q_\alpha]}{\dd Q_\alpha},\quad \frac{Q_\alpha}{\hat{g}_{\alpha} } =\frac {\dd\ln z[Q_\alpha^*]}{\dd Q^*_\alpha}.
\end{equation}
The application of the saddle point method gives
\begin{multline}\label{Znsaddle}
\overline{Z^n}\simeq\exp\left(-NS[Q^\text{sp}]-N\Delta S^T[Q^\text{sp}]-\frac{1}{2}\ln \det\boldsymbol \Omega[Q^\text{sp}]\right),
\end{multline}
where $\boldsymbol \Omega$ is the Hessian matrix of $S[Q]$ and $Q^\text{sp}$ is the saddle-point solution. As we will show below, the contribution $\ln \det\boldsymbol \Omega[Q^\text{sp}]$ provides finite-size corrections to the leading contribution of the same order of the corrections in $N\Delta S^T[Q^\text{sp}]$.
\subsection{Replica symmetric ansatz and limit of vanishing number of replicas}
To proceed with our calculation, we adopt, as usual in the literature, a replica symmetric ansatz for the solution of the saddle-point equations. A replica symmetric solution is of the form
\begin{equation}
Q_\alpha = Q^*_\alpha =q_{|\alpha|}. \label{rsa}
\end{equation}
In particular this implies that $Y_\alpha = 0$. In order to analytically continue to $n\to 0$ the value at the saddle-point of $S$ in Eq.~\eqref{S}, let us first remark that under the assumption in Eq.~\eqref{rsa}
\begin{equation}
\sideset{}{'} \sum_{\alpha\in {\cal P}([n])}  \frac{|Q_\alpha|^2}{ \hat{g}_{\alpha} }= 
\sum_{k=1}^n \binom{n}{k} \frac{q_k^2}{ \hat{g}_k }
=n\sum_{k=1}^\infty\frac{(-1)^{k-1}}{k} \frac{q_k^2}{\hat{g}_k}+o(n).
\end{equation}
Moreover, as shown in Appendix \ref{app:zq},
\begin{equation}
\sideset{}{'} \sum_{\alpha\in {\cal P}([n])} \ln z[Q_\alpha] 
= n\int_{-\infty}^{+\infty}\left(\e^{-\e^l}-\e^{-G(l)}\right)\dd l,
\end{equation}
where
\begin{equation}
G(l)\coloneqq \sum_{k=1}^\infty (-1)^{k-1} q_k \frac{\e^{lk}}{k!}.
\end{equation}
In conclusion, under the replica symmetric ansatz in Eq.~\eqref{rsa}, the functional to be minimized is
\begin{equation}
\hat{\beta} \hat{F} = \sum_{k=1}^\infty\frac{(-1)^{k-1}}{k}\frac{q_k^2}{\hat{g}_k}-2\int_{-\infty}^{+\infty}  \left[\e^{-\e^l} -\e^{-G(l)}\right]\dd l.
\end{equation}
A variation with respect to $q_k$ gives the saddle-point equation
\begin{equation}\label{saddle1}
\frac{1}{k}\frac{q_k}{\hat{g}_k}=\int_{-\infty}^{+\infty}\e^{-G(y)}\frac{\e^{y k}}{k!}\dd y 
\end{equation}
which is to say
\begin{multline}
G(l) = \sum_{k=1}^\infty (-1)^{k-1} q_k \frac{\e^{lk}}{k!}\\ = \int_{-\infty}^{+\infty}\e^{-G(y)}
\sum_{k=1}^\infty (-1)^{k-1} k\hat{g}_k\frac{\e^{(y+l)\, k}}{(k!)^2}\dd y. \label{43}
\end{multline}
This implies that
\begin{multline}
\sum_{k=1}^\infty \frac{(-1)^{k-1}}{k} \frac{q_k^2}{\hat{g}_k} =
\sum_{k= 1}^\infty (-1)^{k-1} q_k \int_{-\infty}^{+\infty} \e^{-G(y)}  \frac{\e^{yk}}{k!}\dd y\\
= \int_{-\infty}^{+\infty}G(y)\e^{-G(y)}\dd y.
\end{multline}
These formulas are for a general law $\rho_r$. Observe also that the expression of $\hat g_p$ is not specified. For finite $r$ and $N\to\infty$, Eq.~\eqref{ghat} simplifies as
\begin{equation}\label{ghat2}
\lim_{N\to\infty} \hat{g}_p = \frac{\eta_0\Gamma(r+1)}{(\hat \beta p)^{r+1}}.
\end{equation} 
We will restrict the analysis to the case in which Eq.~\eqref{ghat2} holds. Then Eq.~\eqref{43} becomes
\begin{equation}
G_r(l) = \frac{\eta_0 \Gamma(r+1)}{\hat{\beta}^{r+1}} \int_{-\infty}^{+\infty} B_{r}(l+y)\e^{-G_r(y)}\dd y,\label{G}
\end{equation}
with
\begin{equation}
B_r(x) \coloneqq\sum_{k=1}^\infty (-1)^{k-1}\frac{\e^{x k}}{k^r(k!)^2}. \label{Br}
\end{equation}
In Eq.~\eqref{G}, and in the following, we introduce the subindex $r$ to stress the dependence of $G$ and of the thermodynamical functionals on $r$. The average cost is therefore
\begin{equation}
\hat E_r=\frac{\partial}{\partial\hat\beta}\hat\beta \hat F_r=\frac{r+1}{\hat\beta}\int_{-\infty}^{+\infty}G_r(y)\e^{-G_r(y)}\dd y.
\end{equation}
Using the fact that (see Appendix \ref{app:B})
\begin{equation}
\lim_{\delta\to\infty} \frac{1}{\delta^r} B_r(\delta x) =\frac{x^r\theta(x)}{\Gamma(r+1)},
\end{equation}
if we introduce
\begin{equation}
\hat{G}_r(l)\coloneqq G_r\left (\frac{\hat{\beta}}{[\eta_0\Gamma(r+1)]^\frac{1}{r+1}}\, l\right)\label{Grdef}
\end{equation}
in the limit $\hat{\beta} \to +\infty$, the function $\hat{G}_r$ satisfies Eq.~\eqref{Gr}
and the value of $\hat{E}_r$ is the one reported in Eq.~\eqref{Er}.
In particular, at fixed $r$, if we consider the two laws $\rho_r^\text{P}$ and $\rho_r^\Gamma$, the ratio between the corresponding average optimal costs is given by
\begin{equation}
\lambda_r\coloneqq\frac{ \hat{E}^\text{P}_r }{ \hat{E}^\Gamma_r } =  \left(\frac{\eta_0^\Gamma}{\eta_0^\text{P}}\right)^\frac{1}{r+1} =[\Gamma(r+2)]^{-\frac{1}{r+1}}.\label{ratio}
\end{equation}
In the case $r=0$, we have the classical result by \textcite{Mezard1985}
\begin{align}
\hat{G}_0(l)&= \ln (1 + \e^l),\\
\hat{E}_0&= \frac{1}{\eta_0(0)}\int_{-\infty}^{+\infty} \frac{\ln (1 + \e^y)}{1 + \e^y} \dd y=  \frac{1}{\eta_0(0)}\frac{\pi^2}{6},
\end{align}
a result that was later obtained with a cavity approach by \textcite{Aldous2001}. For the evaluation of the integral, see Appendix~\ref{app:J}.

\section{Finite size corrections}\label{sec:fsc}
The evaluation of the first-order corrections for a finite number of points has been considered in Refs.~\cite{Mezard1987,Ratieville2002} in the $r=0$ case. For this particular choice and assuming a distribution law $\rho^\Gamma_0$, a much stronger conjecture was proposed by \textcite{Parisi1998} and later proved by \textcite{Linusson2004,Nair2005}, that is, for every $N$,
\begin{equation}\label{Pconj}
\hat{E}_0^\Gamma(N)= H_{N,2}  \coloneqq\sum_{k=1}^N \frac{1}{k^2}.
\end{equation}
For large $N$, Parisi's formula implies
\begin{equation}
\hat{E}_0^\Gamma(N)= \frac{\pi^2}{6} - \frac{1}{N}+o\left(\frac{1}{N}\right).\label{PC}
\end{equation}
Using instead the law $\rho_0^\text{P}$ (uniform distribution on the interval) we have \cite{Mezard1987,Ratieville2002} 
\begin{equation}
\hat{E}_0^\text{P}(N) =\frac{\pi^2}{6} - \frac{1 +  2\zeta(3)}{N}+o\left(\frac{1}{N}\right)\label{72}
\end{equation}
from which we see that  corrections for both laws are analytic, with the same inverse power of $N$, but different coefficients.

In their study of the finite-size corrections, the authors of Ref.~\cite{Ratieville2002} show that, in their particular case, there are two kind of finite-size corrections. The first one comes from the  application of the saddle-point method, giving a series of corrections in the inverse powers of $N$. This contribution is the sum of two terms. The first term in this expansion corresponds to the contribution of the $\Delta S^T$ term given in Eq.~\eqref{DST} appearing in the exponent in Eq.~\eqref{Znsaddle}. The second term is related to the fluctuations, also appearing in Eq.~\eqref{Znsaddle}, involving the Hessian of $S$. The second kind of corrections, instead, is due to the particular form of the law $\rho_r(w)$ for the random links and in particular to the series expansion in Eq.~\eqref{ghat}. This contribution can be seen at the level of the action $S$ in Eq.~\eqref{S}, being
\begin{multline}\label{sommak}
\frac{|Q_\alpha|^2}{\hat{g}_{\alpha} }\approx |Q_\alpha|^2\frac{(\hat \beta |\alpha|)^{r+1}}{\eta_0\Gamma(r+1)}\\
-|Q_\alpha|^2 \frac{(r+1)}{N^{\frac{1}{r+1}}}\frac{\eta_1  }{\eta^2_0} \frac{(\hat \beta |\alpha|)^{r}}{\Gamma(r+1)} + O\left(N^{-\frac{2}{r+1}}\right).
\end{multline}
In full generality, the expansion of $1/\hat{g}_{\alpha}$ generates a sum over terms each one of order $N^{-\frac{k}{r+1}}$ with $k\geq 1$. All these corrections are $o(N^{-1})$ for $r\in(-1,0)$, whereas the corrections obtained from the contributions with $1\leq k\leq r+1$ are of the same order as the analytic term, or greater, for $r\geq 0$. In particular, if $\eta_1\neq 0$, for $r> 0$ the $k=1$ term provides the leading correction, scaling as $N^{-\frac{1}{r+1}}$. It is also evident that all these corrections are absent if $\eta_k=0$ for $k\geq 1$, as it happens in the case of the $\rho_r^\text{P}$ law.

\subsection{Correction due to $\eta_1$}
Let us consider the $r\geq 0$ case and let us restrict ourselves to the $k=1$ term, of order $N^{-\frac{1}{r+1}}$ in Eq.~\eqref{resume}. Its contribution to the total free energy is given by 
\begin{equation}
\hat \beta\Delta\hat F_r^{(1)}
=-\frac{r+1}{N^{\frac{1}{r+1}}}\frac{\eta_1}{\eta_0^2}\sum_{p=1}^\infty\frac{(-1)^{p-1}}{p} \frac{(\hat \beta p)^{r}}{\Gamma(r+1)}q_p^2,
\end{equation}
where we already made a replica symmetric assumption and considered the $n\to 0$ limit. Imposing the saddle-point relation in Eq.~\eqref{saddle1} and using the limit in Eq.~\eqref{ghat2}, we obtain
\begin{multline}
N^{\frac{1}{r+1}}\Delta\hat F_r^{(1)}=\\=-\frac{\eta_1(r+1)}{\hat \beta^{2}\eta_0}\int_{-\infty}^{+\infty}\e^{-G_r(y)}\sum_{p=1}^\infty \frac{(-1)^{p-1} q_p \e^{py}}{p \, p!}\dd y\\
=-\frac{\eta_1(r+1)}{\hat \beta^{2}\eta_0}\int_{-\infty}^{+\infty} \e^{-G_r(y)}\int_{-\infty}^y G_r(u)\dd y\dd u\\
=-\frac{\eta_1(r+1)}{\eta_0\left[\eta_0\Gamma(r+1)\right]^\frac{2}{r+1}}\int_{-\infty}^{+\infty} \e^{-\hat G_r(y)}\int_{-\infty}^y \hat G_r(u)\dd y\dd u.
\end{multline}
To put the expression above in the form presented in Eq.~\eqref{DF1}, observe that
\begin{multline}
\int_{-\infty}^{+\infty} \e^{-\hat G_r(y)}\int_{-\infty}^y \hat G_r(u)\dd y\dd u=\\
=\int_{-\infty}^{+\infty} \hat G_r(-u) \int_{-\infty}^{+\infty}\e^{-\hat G_r(y)} \theta(y+u)\dd y\dd u\\
=\int_{-\infty}^{+\infty} \hat G_r(-u) \fder{r}{u}\hat G_r(u)\dd u\equiv J_r^{(r)},
\end{multline}
a structure that can be more useful for numerical evaluation, at least for $r$ integer. In this equation we have used Eq.~\eqref{intexpg} and we have introduced the Riemann--Liouville fractional derivative
\begin{multline}
\fder{\alpha}{t}f(t)\coloneqq\frac{\dd^{[\alpha]+1}}{\dd t^{[\alpha]+1}}\int_{-\infty}^t\frac{(t-\tau)^{[\alpha]-\alpha}}{\Gamma([\alpha]-\alpha+1)}f(\tau)\dd\tau,\\
\alpha\geq 0,\quad f\in L_p(\Omega)\ \forall p\in\left[1,\frac{1}{[\alpha]-\alpha+1}\right),
\end{multline}
where $\Omega\coloneqq(-\infty,t)$ is the domain of integration (see the monographs in Refs.~\cite{Samko1993,Podlubny1999} for further details).
\subsection{Correction due to the saddle point approximation}
Let us now consider the corrections due to the saddle point approximation. The first contribution is expressed by $\Delta S^T$, given in Eq.~\eqref{DST}. In the replica symmetric hypothesis, we have that
\begin{multline}
\sideset{}{'}\sum_{\mathclap{\substack{\alpha,\beta\in {\cal P}([n])\\\alpha\cap\beta=\emptyset}}}\hat g_\alpha\hat g_\beta\frac{|Q_{\alpha\cup\beta}|^2}{2N\hat{g}^2_{\alpha\cup\beta}}=\\= \frac{1}{2N}\sum_{s=1}^\infty \sum_{t=1}^\infty \binom{n}{s, t, n-s-t}\frac{\hat{g}_s \hat{g}_t}{\hat{g}^2_{s+t} }q_{s+t}^2.
\end{multline}
We can write the corresponding correction to the free energy as
\begin{equation}\label{DFT}
\Delta \hat{F}_r^T = \frac{1}{2 \hat{\beta} N} \sum_{s=1}^\infty \sum_{t=1}^\infty (-1)^{s+t -1} \frac{(s+t-1)!}{s! \,t!} \frac{\hat{g}_s \hat{g}_t}{\hat{g}^2_{s+t} } q_{s+t}^2.
\end{equation}
In Appendix \ref{app:FT} we show that the previous quantity can be written as
\begin{multline}\label{DFTfin}
\Delta \hat{F}_r^T=-\frac{\Gamma(2r+2)}{N\eta_0^{\frac{1}{r+1}}\Gamma^{2+\frac{1}{r+1}}(r+1)}\frac{1}{r+1}\int_{-\infty}^{+\infty}\hat G_r(-u)\hat G_r(u)\dd u\\
=-\frac{\Gamma(2r+2)}{N\eta_0^{\frac{1}{r+1}}\Gamma^{2+\frac{1}{r+1}}(r+1)}\frac{J_r^{(0)}}{r+1}.
\end{multline}

Another type of finite-size correction comes from the fluctuations around the saddle point \cite[Section B.3]{Ratieville2002}, related to the Hessian matrix $\boldsymbol\Omega$ appearing in Eq.~\eqref{Znsaddle}. The evaluation of the contribution of the Hessian matrix is not trivial and it has been discussed by \textcite{Mezard1987} and later by \textcite{Ratieville2002}. They proved that the whole contribution comes from a volume factor due to a non trivial metric $\hat{\boldsymbol \Omega}$ obtained from $\boldsymbol \Omega$ imposing the replica symmetric assumption and such that
\begin{equation}
\ln\sqrt{\det{\boldsymbol\Omega}}=\ln\sqrt{\det\hat{\boldsymbol\Omega}}.
\end{equation}
The $n\times n$ matrix $\hat{\boldsymbol \Omega}$ can be written as
\begin{equation}
\hat{\boldsymbol\Omega}=na_1\boldsymbol\Pi+  (a_0-a_1)\mathbb{I}_n,
\end{equation}
where $\mathbb I_n$ is the $n\times n$ identity matrix and we have introduced the quantities
\begin{subequations}
\begin{align}
a_0&\coloneqq\sum_{p=1}^\infty {n-1 \choose p-1}\frac{q_p^2}{\hat{g}_p},\\
a_1&\coloneqq\sum_{p=2}^\infty {n-2 \choose p-2}\frac{q_p^2}{\hat{g}_p},
\end{align}
\end{subequations}
and $\boldsymbol{\Pi}$ is a projection matrix on the constant vector defined as
\begin{equation}
\boldsymbol{\Pi}\coloneqq\frac{\mathbb{J}_n}{n},
\end{equation}
where $\mathbb{J}_n$ is the $n\times n$ matrix with all entries equal to $1$. The matrix $\boldsymbol\Pi$ has one eigenvalue equal to $1$ and $n-1$ eigenvalues equal to $0$. It follows that, because the two matrices $\boldsymbol\Pi$ and $\mathbb I_n$ obviously commute, $\hat{\boldsymbol\Omega}$ has one eigenvalue equal to $a_0+(n-1)a_1$ and $n-1$ eigenvalues equal to $a_0-a_1$. Its determinant is therefore simply given by
\begin{equation}
\det\hat{\boldsymbol\Omega}=(a_0-a_1)^{n-1}[a_0 + (n-1) a_1].
\end{equation}
In the limit of $n\to 0$ we easily get
\begin{multline}
a_0=\sum_{p=1}^\infty (-1)^{p-1}\frac{q_p^2}{\hat{g}_p}
=\sum_{p=1}^\infty (-1)^{p-1} pq_p\int_{-\infty}^{+\infty} \e^{-G_r(y)}\frac{\e^{py}}{p!}\dd y\\
=\int_{-\infty}^{+\infty}\e^{-G_r(y)} \frac{\dd G_r(y)}{\dd y}\dd y=\int_{-\infty}^{+\infty} \e^{-\hat{G}_r(y)} \frac{\dd\hat{G}_r(y)}{\dd y} \dd y  \\
= -\int_{-\infty}^{+\infty}\frac{\dd}{\dd y} \e^{-\hat{G}_r(y)}\dd y=\e^{-\hat{G}_r(-\infty)} - \e^{-\hat{G}_r(+\infty)}=1
\end{multline}
for all values of $r$. Similarly,
\begin{multline}
a_1 =-\sum_{p=2}^\infty (-1)^{p-1}(p-1 )\frac{q_p^2}{\hat{g}_p} = -\sum_{p= 1}^\infty (-1)^{p-1}(p-1 )\frac{q_p^2}{\hat{g}_p}
\end{multline}
so
\begin{multline}
a_0 - a_1 =\sum_{p= 1}^\infty (-1)^{p-1}p\frac{q_p^2}{\hat{g}_p}\\
= \sum_{p= 1}^\infty (-1)^{p-1}p^2q_p\int_{-\infty}^{+\infty}\e^{-G_r(y)}\frac{\e^{py}}{p!}\dd y \\
= \int_{-\infty}^{+\infty} \e^{-G_r(y)}\frac{\dd^2}{\dd y^2} G_r(y)\dd y\\
= \frac{\left[\eta_0\Gamma(r+1) \right]^\frac{1}{r+1}}{\hat{\beta}} \int_{-\infty}^{+\infty}\e^{-\hat{G}_r(y)} \frac{\dd^2}{\dd y^2}  \hat{G}_r(y)  \dd y.
\end{multline}
Therefore,
\begin{equation}
\sqrt{\det \hat{\boldsymbol\Omega}} = 1+\frac{n}{2}\left[\frac{a_1}{a_0-a_1}+\ln(a_0-a_1)\right]+o(n).
\end{equation}
In conclusion, integrating by parts and using Eq.~\eqref{Grderr}, we obtain
\begin{multline}
\Delta \hat{F}_r^F =-\lim_{\hat{\beta} \to \infty} \frac{1}{n N \hat{\beta}} \ln \sqrt{\det \hat{\boldsymbol\Omega}}\\
=- \frac{1}{2N\left[\eta_0\Gamma(r+1)\right]^\frac{1}{r+1} } \frac{1}{J_{r}^{(r+3)}}.
\end{multline}

\subsection{Application: the $r=0$ case}
The results obtained in the $r=0$ case, analyzed by \textcite{Ratieville2002}, can be easily recovered. From the general expression in Eq.~\eqref{resume}, by setting $r=0$, we get
\begin{multline}
\Delta \hat{F}_0\coloneqq\Delta \hat{F}_0^{(1)}+\Delta \hat{F}_0^T+\Delta \hat{F}_0^F
\\= - \frac{1}{\eta_0(0) N} \left[\left(1+\frac{\eta_1(0)}{\eta_0(0)}\right)\frac{J_0^{(0)}}{\eta_0(0)} +\frac{1}{2J_0^{(3)}} \right]\\
=- \frac{1}{\eta_0(0) N} \left[\left(1+\frac{\eta_1(0)}{\eta_0(0)}\right)\frac{2\zeta(3)}{\eta_0(0)}+1 \right],
\end{multline}
where we have used the results discussed in the Appendix \ref{app:J} for the two integrals involved in the expression above. Eqs.~\eqref{PC} and \eqref{72} are obtained using Eqs.~\eqref{cGamma} and \eqref{cP}, respectively.

\section{The limiting case $r\to +\infty$}\label{sec:rinf}
In this section we concentrate on the limiting case in which $r\to +\infty$. We can easily verify that, in the weak sense,
\begin{equation}
\lim_{r \to + \infty} \rho_r^\text{P}(w) = \delta(w-1)
\end{equation}
so all the weights become equal to unity. We expect therefore that
\begin{equation}
\lim_{r \to + \infty} \hat{E}_r^\text{P}(N) = 1,
\end{equation}
independently of $N$. The average cost obtained using $\rho^\Gamma_r$ instead diverges and it is therefore more interesting to consider the modified law
\begin{equation}
\rho_r^{\gamma}(w)\coloneqq \frac{(r+1)^{r+1}}{\Gamma(r+1)}w^r \e^{-(r+1) w}\theta(w)\xrightarrow{r\to\infty}\delta(w-1).
\end{equation}
According to our general discussion, we have that
\begin{equation}
\eta^\gamma_k(r)=\frac{(r+1)^{k+r+1}}{\Gamma(r+1)}\frac{(-1)^k}{k!},\quad k\geq 0,\label{cgamma}
\end{equation}
implying that, independently of $N$,
\begin{equation}
\hat{E}_r^{\gamma}(N)  = \frac{1}{r+1}\hat{E}_r^\Gamma(N)
\end{equation}
and therefore
\begin{equation}
\hat{E}_r^{\gamma}=\frac{\Gamma(r+2)^\frac{1}{r+1}}{r+1} \hat{E}_r^\text{P}
\end{equation}
in the limit of infinite $N$. In particular,
\begin{equation}
\lim_{r \to + \infty} \hat{E}_r^{\gamma}  = \frac{1}{\e}.
\end{equation}
It follows that, even though the two laws $\rho^\text{P}$ and $\rho^\gamma$ both converge to the same limiting distribution, according to our formulas, the corresponding average costs are \textit{not} the same. This is due to the fact that the two limits $N\to +\infty$ and $r\to +\infty$ do not commute for the law $\rho^\gamma$, because of the presence of $O(N^{-\frac{k}{r+1}})$ corrections that give a leading contribution if the $r\to\infty$ limit is taken first.

To look into more details in the $r\to +\infty$ limit, we find it convenient, when looking at the saddle-point solution, to perform a change of variables, following the approach in Refs.~\cite{Houdayer1998,Parisi2001}, that is, writing
\begin{equation}
\mG_r(x)\coloneqq \hat G_r\left[\Gamma^\frac{1}{r+1}(r+2)\left( \frac{1}{2} + \frac{x}{r+1}\right) \right]
\label{cv}\end{equation}
then Eq.~\eqref{Gr} becomes
\begin{equation}
\mG_r(x) = \int_{-x-r-1}^{+\infty}\left(1 + \frac{x+t}{r+1}\right)^r\e^{-\mG_r(t)}\dd t,
\end{equation}
so, in the $r \to +\infty$ limit
\begin{equation}
\mG_\infty(x) = \e^x \int_{-\infty}^{+\infty}\e^{t-\mG_\infty(t)}\dd t.
\end{equation}
If we set $\mG_\infty(x) = a \e^x$, with
\begin{equation}
a = \int_{-\infty}^{+\infty}\e^{t-\mG_\infty(t)}\dd t
\end{equation}
we recover
\begin{equation}
a = \int_{-\infty}^{+\infty}\e^{t- a \e^t}\dd t = \int_0^{+\infty} \e^{-a z}\dd z= \frac{1}{a}\Rightarrow \mG_\infty(x)=\e^x.
\end{equation}
From Eq.~\eqref{Er} with the change of variable in Eq.~\eqref{cv}, we get
\begin{equation}
\hat{E}_r = \left( \frac{r+1}{\eta_0(r)} \right)^\frac{1}{r+1}  \int_{-\infty}^{+\infty}\mG_r(x)\e^{-\mG_r(x)} \dd x,
\end{equation}
so
\begin{equation}
\hat{E}_\infty  =  \lim_{r \to + \infty} \left( \frac{r+1}{\eta_0(r)} \right)^\frac{1}{r+1} = 
\lim_{r \to + \infty} {\eta_0(r)}^{-\frac{1}{r+1}}=\begin{cases}
1 & \text{for $\rho^\text{P}_r$,} \\
\frac{1}{\e} & \text{for $\rho^\gamma_r$,}
\end{cases}
\end{equation}
in agreement with the previous results.

Let us now evaluate the integrals appearing in the finite-size corrections in Eq.~\eqref{resume}. Let us first start with the $\Delta F^T_r$ and the $\Delta F_r^F$ corrections. From the definition, for large $r$,
\begin{multline}
J_r^{(0)} =\int_{-\infty}^{+\infty}\hat{G}_r(y)\hat{G}_r(- y)\dd y\\
=\int_{-\infty}^{+\infty} \dd y  \int_{-y}^{\infty} \dd t_1\frac{(t_1+y)^r}{\Gamma(r+1)} \e^{- \hat{G}_r(t_1)}\int_{y}^{\infty} \dd t_2\frac{(t_2+y)^r}{\Gamma(r+1)} \e^{- \hat{G}_r(t_2)}\\
=\Gamma^\frac{1}{r+1}(r+2) \int_{-\infty}^{+\infty}\dd t_1  \int_{-\infty}^{+\infty}\dd t_2\, K_r(t_1,t_2)\e^{- \mG_r(t_1) - \mG_r(t_2)}
\end{multline}
where
\begin{multline}
K_r(t_1,t_2)\coloneqq \int_{-x_1-r-1}^{x_2} \frac{(t+t_1+r+1)^r(t_2-t)^r}{(r+1)^{2r+1}}\dd t\\
=\left(1+ \frac{x_1 + x_2}{r+1}\right)^{2r+1} \frac{\Gamma^2(r+1)}{\Gamma(2r+2)} 
\end{multline}
and therefore for large $r$,
\begin{multline}
J_r^{(0)} 
\simeq \Gamma^\frac{1}{r+1}(r+2)\frac{\Gamma^2(r+1)}{\Gamma(2r+2)}\left( \int_{-\infty}^{+\infty} \e^{x-\e^x}\dd x\right)^2 \\
= \Gamma^\frac{1}{r+1}(r+2)\,  \frac{\Gamma^2(r+1)}{\Gamma(2r+2)}\simeq \frac{\Gamma^{2+\frac{1}{r+1}}(r+1)}{\Gamma(2r+2)},
\end{multline}
so we get 
\begin{equation}
\Delta \hat{F}_r^T \simeq  - \frac{1}{N \eta_0^\frac{1}{r+1}(r) r} 
\end{equation}
a contribution that vanishes as $r^{-1}$ both for the law $\rho^\text{P}_r$ and for the law $\rho^\gamma_r$ (indeed we know that, in this case, all corrections must vanish when $r\to \infty$ at fixed $N$). In particular, if we consider the law $\rho^\Gamma_r$, we have, in the $r\to\infty$ limit,
\begin{equation}
\Delta \hat{F}_\infty^T=-\frac{1}{\e N}.
\end{equation}
Similarly, for large $r$, we have that
\begin{equation}
\frac{1}{J^{(r+3)}_r}=\frac{\Gamma(r+2)^\frac{1}{r+1}}{r+1}\left[\int_{-\infty}^{+\infty}\e^{- \mG_r(x)} \left( \frac{\dd}{\dd x} \right)^2 \mG_r(x)\right]^{-1}=\frac{1}{\e}
\end{equation}
and therefore
\begin{equation}
\Delta \hat{F}_r^F \simeq  - \frac{1}{2 N \eta_0^\frac{1}{r+1}(r) r}=\frac{\Delta \hat{F}_r^T}{2}.
\end{equation}

Instead, if we consider $\Delta \hat F^{(1)}_r$, we have that
\begin{multline}
J_r^{(r)} = \left[\frac{\Gamma(r+2)^\frac{1}{r+1}}{r+1} \right]^2 \int_{-\infty}^{+\infty}\dd u\, \e^{-\mG_r(u)} \int_{-\infty}^u \dd v \, \mG_r(v) \\
\xrightarrow{r\to\infty} \frac{1}{\e^2} \int_{-\infty}^{+\infty} \dd u \, \e^{-\e^u} \int_{-\infty}^u \dd v \, \e^v=\frac{1}{ \e^2},
\end{multline}
finally obtaining
\begin{equation}
\Delta \hat{F}_r^{(1)} \simeq - \frac{\eta_1(r)}{N^\frac{1}{r+1}\eta_0^{1+\frac{2}{r+1}}(r) r}
\end{equation}
so that, considering the law $\rho^\gamma_r$, if we send $r\to \infty$ \textit{before} taking the limit $N \to \infty$, $\Delta \hat{F}_{+\infty}^{(1)}\sim O(1)$ and we get a new contribution to the average optimal cost
\begin{equation}
\hat{E} = \hat{E}_\infty +\sum_{k=1}^\infty \Delta \hat{F}_\infty^{(k)}= \frac{1}{\e} + \frac{1}{\e^2} + \dots
\end{equation}
a series where we miss the contributions of order $N^\frac{k}{r+1}$ for $k\geq 2$, and that we know will sum to $1$.

\section{Numerical results}\label{sec:num}
In this section we discuss some numerical results. First, we present a numerical study of our theoretical predictions obtained in the previous sections. Second, we compare with numerical simulations, in which the random assignment problem is solved using an exact algorithm.

The evaluation of all quantities in Eq.~\eqref{resume} depends on the solution of Eq.~\eqref{Gr}. We solved numerically this equation for general $r$ by a simple iterative procedure. In particular, for $r >0$ we generated a grid of $2K-1$ equispaced points in an interval $\left[ -y_{\text{max}},  y_{\text{max}} \right]$ and we used a discretized version of the saddle-point equation in Eq.~\eqref{Gr} in the form
\begin{multline}\label{discretizzata}
\hat G_r^{[s+1]} (y_i)=\frac{y_\text{max}}{K}\sum_{k=2K-i}^{2K}\frac{(y_i+y_k)^r \e^{-\hat G_r^{[s]}(y_k)}}{\Gamma(r+1)},\\
y_i=\frac{i-K}{K}y_\text{max},\quad i=0,1,\dots,2K.
\end{multline}
We imposed as the initial function $\hat G_r^{[0]}$ of the iterative procedure
\begin{equation}
\hat G_r^{[0]} (y_i) \equiv \hat G_0(y_i) = \ln \left( 1 + \e^{y_i} \right).
\end{equation}
We observed that the quantity
\begin{equation}
\Delta G_r^{[s]} = \sum_{i=0}^{2K} \left|\hat G_r^{[s]} (y_i) - \hat G_r^{[s-1]} (y_i)  \right|
\end{equation}
decays exponentially with $s$ and therefore convergence is very fast. For our computation, we used typically $30$ iterations. 

For $r<0$ the term $(l+y)^r$ in the saddle-point equation is divergent in $y=-l$ and Eq.~\eqref{discretizzata} cannot be adopted. We have therefore rewritten the saddle-point equation using an integration by parts, obtaining
\begin{equation}
\hat G_r(l) = \int_{-l}^{+\infty}\frac{\left( l+y \right)^{r+1}\e^{-\hat G_r(y)}}{\Gamma(r+2)} \frac{\dd \hat G_r(y)}{\dd y}\dd y.
\end{equation}
After discretizing the previous equation, we used the same algorithm described for the $r \ge 0$ case (for a discussion on the uniqueness of the solution of Eq.~\eqref{Gr}, see Ref.~\cite{Salez2015}). 

In Table~\ref{tab:J} we present our numerical results for the quantities 
\[\eta_0^{\frac{1}{r+1}} \hat E_r,\quad (N \eta_0^2)^{\frac{1}{r+1}} \frac{\eta_0\Delta \hat F^{(1)}_r}{\eta_1},\quad N \eta_0^{\frac{1}{r+1}} \Delta\hat F^T_r,\quad N \eta_0^{\frac{1}{r+1}}  \Delta \hat F^F_r\] 
for different values of $r$. Observe that the quantities appearing in the expansion in Eqs.~\eqref{resume} can be calculated using these values for any $\rho_r$ at given $r$, in addition to simple prefactors depending on the chosen distribution $\rho_r$.

\begin{table}\squeezetable
\begin{center}
\begin{ruledtabular}
\begin{tabular}[b]{ccccc}
$r$ & $\eta_0^{\frac{1}{r+1}} \hat E_r$ & $ (N \eta_0^2)^{\frac{1}{r+1}} \frac{\eta_0\Delta\hat  F^{(1)}_r}{\eta_1} $ & $ N \eta_0^{\frac{1}{r+1}} \Delta \hat F^T_r $ & $ N \eta_0^{\frac{1}{r+1}}  \Delta \hat F^F_r $  \\ 
\hline
-0.5   &    1.125775489   &     -2.777285153  &      -3.917446075  &      -1.192663973\\
-0.4    &   1.334614017  &      -2.952484269  &      -3.665262242 &       -1.250475151\\
-0.3    &   1.471169704  &      -2.921791666  &      -3.324960744 &       -1.222990786\\
-0.2      & 1.558280634   &     -2.784084499 &       -2.984917100  &      -1.157857158\\
-0.1    &   1.612502443     &   -2.600804197 &       -2.675513663  &      -1.079610016\\
0    &      1.644934067     &   -2.404113806   &     -2.404113806    &    -1					\\
0.1   &     1.662818967 &       -2.215821874  &      -2.168528577    &    -0.924257491\\
0.2     &   1.671039856  &      -2.038915744 &       -1.966713438   &     -0.854501434\\
0.3  &      1.672729262 &       -1.877696614  &      -1.792481703   &   -0.791231720\\
0.4    &    1.670005231  &      -1.732453452   &     -1.641566768   &     -0.734262435\\
0.5      &  1.664311154   &     -1.602337915   &     -1.510248399    &    -0.683113178\\
0.6  &      1.656639222  &      -1.486024319   &     -1.395391897    &    -0.637204338\\
0.7    &    1.647677145  &      -1.382051819  &      -1.294397704    &    -0.595951390\\
0.8     &   1.637905005  &      -1.288993419  &      -1.205124002  &      -0.558807473\\
0.9     &   1.627659755  &      -1.205532353  &      -1.125808312    &    -0.525279810\\
1       &   1.617178636   &     -1.130489992   &     -1.054997763      &  -0.494933215\\
2    &      1.519733739   &     -0.670341811   &     -0.626403698   &     -0.303146650\\
3    &      1.446919560   &     -0.461144035   &     -0.431759755   &     -0.211631545\\
4    &      1.393163419   &     -0.346056113   &     -0.324185048   &     -0.159938240\\
5    &      1.352087648   &     -0.274505368   &     -0.257174804   &     -0.127356338\\
6    &      1.319651066   &     -0.226200326    &    -0.211931870    &    -0.105200594\\
7    &      1.293333076   &     -0.191617643    &    -0.179566694   &     -0.089276830\\
8    &      1.271505390   &     -0.165752490    &    -0.155385461    &    -0.077340947\\
9    &      1.253073980   &     -0.145742887    &    -0.136697943   &     -0.068095120\\
10  &       1.237277174  &      -0.129842072    &    -0.121861122  &      -0.060741591\\
	\end{tabular}
	\end{ruledtabular}
\end{center}
\caption{Numerical values of the rescaled corrections appearing in Eqs.~\eqref{resume} for different values of $r$.}\label{tab:J}
\end{table}

In order to test our analysis for correction terms, we performed direct Monte Carlo sampling on a set of instances. Previous simulations have been reported, for example, in Refs.~\cite{Mezard1985,Brunetti1991,Houdayer1998,Lee1993}. In our setting, each realization of the matching problem has been solved by a \texttt{C++} implementation of the Jonker-Volgenant algorithm \cite{Jonker1987}. 

We first evaluated the asymptotic average optimal costs $\hat E_r^\text{P}$ and $\hat E_r^\Gamma$, obtained, for different values of $r$, using the laws $\rho_r^\text{P}$ and $\rho_r^\Gamma$, respectively. In the case of the law $\rho_r^\text{P}$, the asymptotic estimate for $\hat E_r^\text{P}$ has been obtained using the fitting function
\begin{equation}\label{fitfunctionsP}
f^\text{P}(N)=\alpha^\text{P}_r+\frac{\beta^\text{P}_r}{N},
\end{equation}
with $\alpha_r^\text{P}$ and $\beta_r^\text{P}$ fitting parameters to be determined, $\alpha_r^\text{P}$ corresponding to the value of the average optimal cost in the $N\to\infty$ limit. For a given value of $r$, we averaged over $I_N$ instances for each value of $N$ accordingly with the table below.
\begin{center}
\begin{ruledtabular}\squeezetable
\begin{tabular}{c|ccccc}
$N$&500&750&1000&2500&5000\\
$I_N$&100000&75000&50000&20000&10000
\end{tabular}
\end{ruledtabular}
\end{center}

\begin{figure}
\centering
\includegraphics[width=\columnwidth]{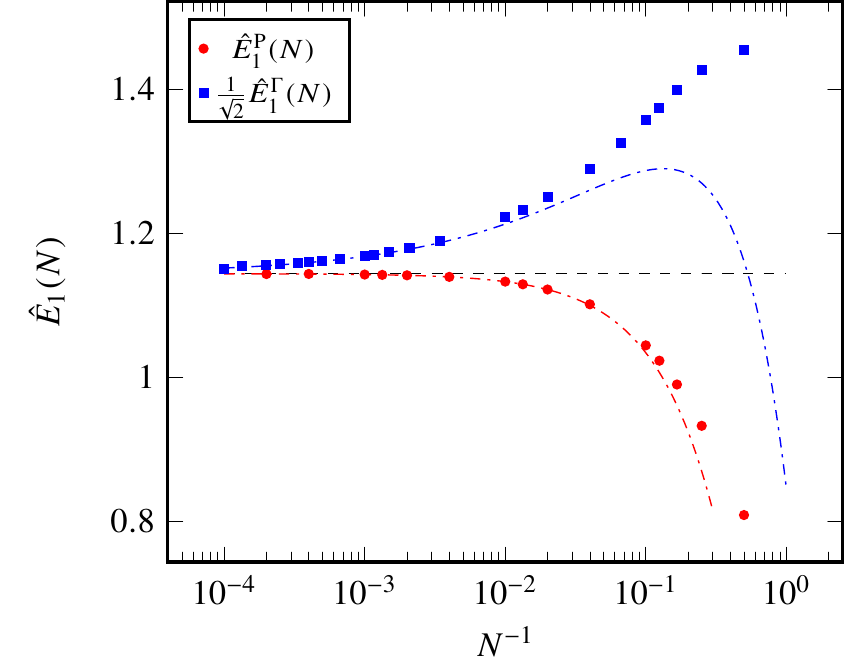}
\caption{Numerical results for $\hat E_1^\text{P}(N)$ and $\hat E_1^\Gamma(N)$ for several values of $N$. Note that finite-size corrections have a different sign for $N\to+\infty$. We have represented also the theoretical predictions for both cases obtained including the finite-size corrections up to $O(N^{-1})$.}\label{fig:r1}
\end{figure}

\begin{figure}
\centering
\includegraphics[width=\columnwidth]{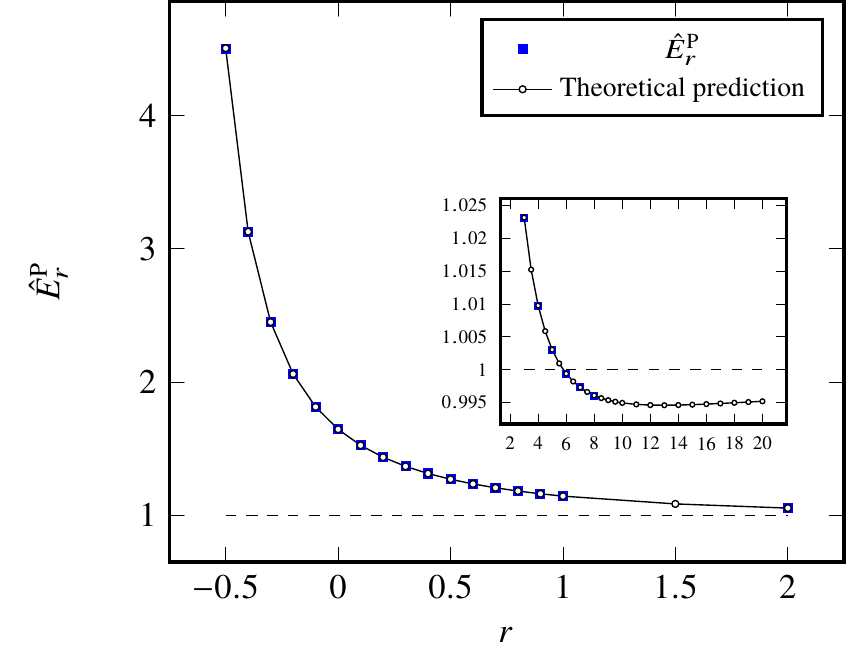}
\caption{Theoretical prediction of $\hat E_r^\text{P}$ for several values of $r$ (solid line), compared with our numerical results. The dashed line is the large-$r$ asymptotic estimate, equal to $1$. Error bars do not appear because they are smaller than the marks in the plot. The values for ${\lambda_r\hat E_r^\Gamma}$ almost coincide with the values of ${\hat E_r^\text{P}}$ (see Table~\ref{tab:costo}) and are not represented.\label{fig:ene}}
\end{figure}

Similarly, the asymptotic average optimal cost $\hat E_r^\Gamma$ has been obtained using a fitting function in the form
\begin{equation}
f^\Gamma (N)=\begin{cases}\alpha^\Gamma_r+\beta^\Gamma_r N^{-1}+ \gamma_r^\Gamma N^{-\frac{1}{r+1}}&\text{for $-\frac{1}{2}\leq r< 1$}\\
\alpha^\Gamma_r+\gamma_r^\Gamma N^{-\frac{1}{r+1}}+\delta_r^\Gamma N^{-\frac{2}{r+1}}&\text{for $r\geq 1$}.\end{cases}\label{fitfunctionsG}
\end{equation}
We adopted therefore a three-parameter fitting function, constructed according to Eq.~\eqref{resume} including the finite-size correction up to $o(N^{-1})$ for $r\geq 0$ and up to $O(N^{-2})$ for $\frac{1}{2}\leq r<2$. As in the case before, the asymptotic estimation for $\hat E_r^\Gamma$ is given by $\alpha^\Gamma_r$. Our data were obtained extrapolating the $N\to\infty$ limit from the average optimal cost  for different values of $N$. The investigated sizes and the number of iterations were the same adopted for the evaluation of $\hat E_r^\text{P}$. To better exemplify the main differences in the finite-size scaling between the $\rho_r^\text{P}$ case and the $\rho_r^\Gamma$ case, we have presented the numerical results for $r=1$ in Fig.~\ref{fig:r1}. In the picture, it is clear that the asymptotic value $\hat E_1^\text{P}=\frac{1}{\sqrt{2}}\hat E_1^\Gamma$ is the same in the two cases, as expected from Eq.~\eqref{ratio}, but the finite-size corrections are different both in sign and in their scaling properties. 
In Table~\ref{tab:costo} we compare the results of our numerical simulations with the ones in the literature (when available) for both $\hat E_r^\text{P}(N)$ and $\lambda_r\hat E_r^\Gamma(N)$, $\lambda_r$ being defined in Eq.~\eqref{ratio}. In Fig.~\ref{fig:ene} we plot our theoretical predictions and the numerical results that are presented in Table~\ref{tab:costo}.
\begin{table}\squeezetable
\begin{center}
\begin{ruledtabular}
\begin{tabular}[b]{cccccc}
$r$ &$\hat E_r^\text{P}$&$\lambda_r \hat E_r^\Gamma$&$\hat E_r^\text{P}$ \cite{Houdayer1998}&$\hat E_r^\text{P}$ \cite{Lee1993}  & Th.~prediction\\ 
\hline
-0.5 & 4.5011(3)&4.504(1)&--&--&4.503101957\\
-0.4 & 3.12611(5)&3.1268(2)&--&--&3.126825159\\
-0.3 & 2.4484(1)&2.4488(3)&--&--&2.448788557\\
-0.2 & 2.0593(5)&2.0593(3)&--&--&2.059601452\\
-0.1 & 1.8127(3)&1.8126(2)&--&--&1.812767212\\
0 & 1.64500(5)&1.6449(2)&1.645(1)&1.6450(1)&1.644934067\\
0.1 & 1.5245(2)&1.5253(9)&--&--&1.524808331\\
0.2 & 1.4356(2)&1.4357(5)&--&--&1.435497487\\
0.3 & 1.3670(1)&1.3670(4)&--&--&1.367026464\\     
0.4 & 1.31323(6)&1.3132(3)&--&--&1.3132296\\
0.5 & 1.27007(8)&1.2697(4)&--&--&1.270107121\\
0.6 & 1.2350(1)&1.2348(3)&--&--&1.234960167\\
0.7 & 1.20585(6)&1.2062(6)&--&--&1.205907312\\
0.8 & 1.18143(3)&1.1812(7)&--&--&1.181600461\\
0.9 & 1.16099(8)&1.1605(6)&--&--&1.161050751\\ 
1 & 1.14344(7)&1.1433(4)&1.143(2)&--&1.14351798\\
2 & 1.05371(1)&1.054(1)&1.054(1)&1.054(1)&1.053724521\\
3 & 1.02311(1)&1.0288(9)&1.0232(1)&1.0236(2)&1.023126632\\
4 & 1.009690(4) & 1.010(3)&1.0098(1)&--&1.009736514\\
5 & 1.00303(2) & 1.005(3)	&1.00306(8)&1.0026(8)&1.003027802\\
\end{tabular}\end{ruledtabular}
\end{center}
\caption{Numerical results for the average optimal cost for different values of $r$ and theoretical predictions. The value $\hat E_r^\text{P}$ from Ref.~\cite{Houdayer1998}, due to a different convention adopted in that paper, is obtained as 
\leavevmode\\\begin{minipage}{\linewidth}\(\hat E_r^\text{P}=\left( \frac{2 \pi^{\frac{r+1}{2}}\Gamma(r+1)}{\Gamma \left(\frac{r+1}{2} \right)\Gamma(r+2)}\right)^{\frac{1}{r+1}}\beta_\text{num}(r+1)\)\end{minipage} from Table~II therein. The data for $\hat E_r^\text{P}$ from Ref.~\cite{Lee1993} have been obtained via a linear fit, using a fitting function in the form of Eq.~\eqref{fitfunctionsP}. }\label{tab:costo}
\end{table}

\begin{figure}
\centering
\includegraphics[width=\columnwidth]{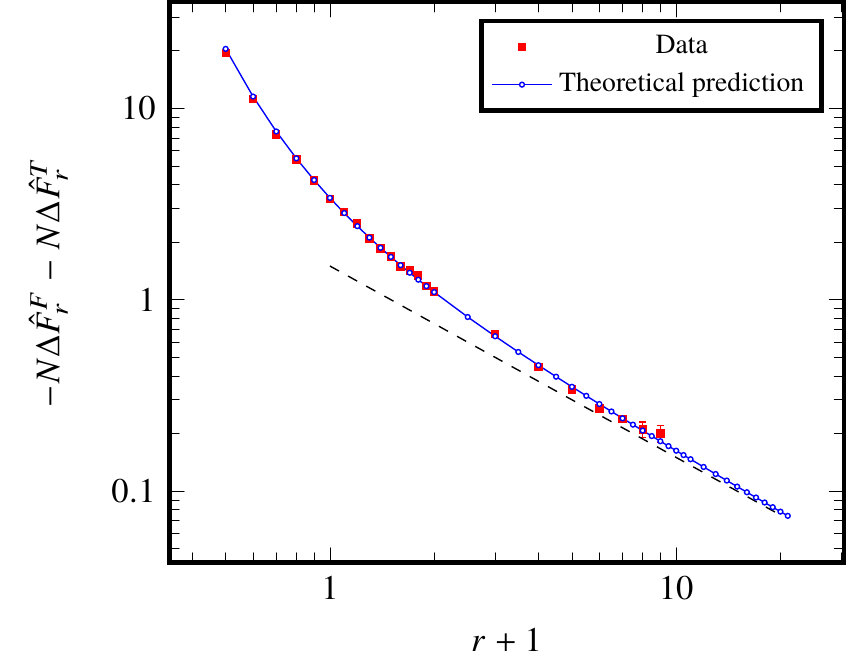}
\caption{Numerical estimates of $\Delta\hat  F_r^T+\Delta\hat  F_r^F$ for several values of $r$ (red squares) and theoretical prediction (blue line) obtained using the law $\rho^\text{P}_r$. The dashed line is the large-$r$ asymptotic estimate.}\label{fig:DFT}
\end{figure}

Let us now consider the finite-size corrections. In the case of the $\rho_r^\text{P}$ law, the $O(N^{-1})$ corrections are given by $\Delta \hat F_r^T+\Delta \hat F_r^F$ and no nonanalytic corrections to the leading term appear. We obtain the finite-size corrections from the data used for Table~\ref{tab:costo}, using Eq.~\eqref{fitfunctionsP} but fixing $\alpha_r^\text{P}$ to the average optimal cost $\hat E_r^\text{P}$ given by the theoretical prediction in Table~\ref{tab:J} and therefore with one free parameter only, namely, $\beta_r^\text{P}$.
In Fig.~\ref{fig:DFT} we compare our predictions for $\Delta \hat F_r^T+\Delta \hat F_r^F$, deduced by the values in Table~\ref{tab:J}, with the results of our numerical simulations for different values of $r$.

\begin{figure}
\centering
\includegraphics[width=\columnwidth]{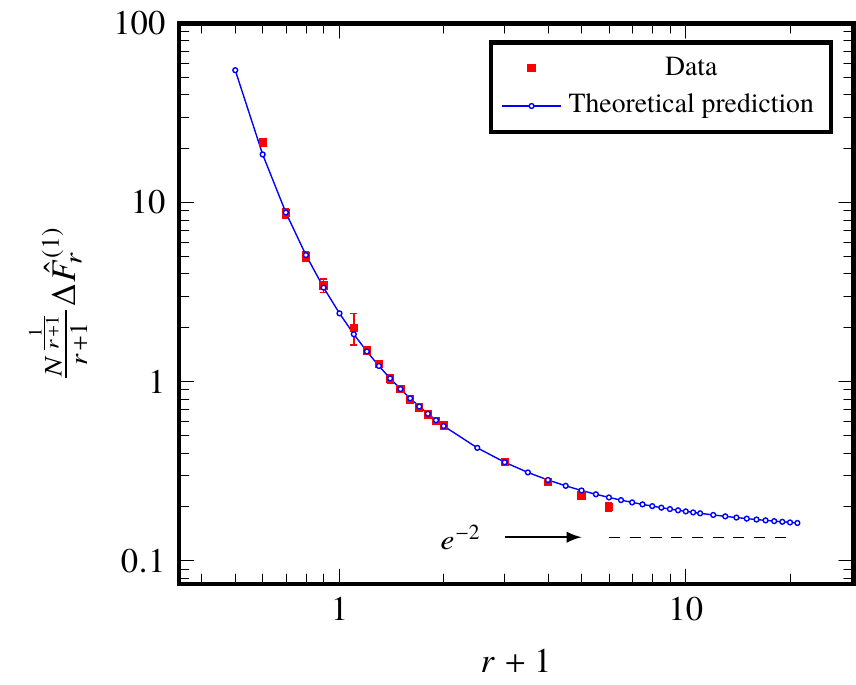}
\caption{Numerical estimates of $\Delta \hat F_r^{(1)}$ for several values of $r$ (red squares) and theoretical prediction (blue line with circles) obtained using the law $\rho^{\Gamma}_r$. Observe that a discrepancy between the theoretical prediction and the numerical results appears for $r\geq 1$: we interpret this fact as a consequence of the similar scaling of $\Delta \hat F^{(1)}_r$ and $\Delta \hat F_r^{(2)}$ for $r\gg 1$, which makes the numerical evaluation of the single contribution $\Delta \hat F^{(1)}$ difficult. The dashed line is the large-$r$ asymptotic estimate.}\label{fig:DFA}
\end{figure}

In the case of the $\rho_r^\Gamma$ law with $r>0$, the first correction to the average optimal cost is given by $\Delta\hat F_r^{(1)}$, whereas $\Delta\hat F_r^{(1)}$ is $o(N^{-1})$ for $r<0$. Again, this correction can be obtained by a fit of the same data used to extrapolate the average optimal cost, fixing the fitting parameter $\alpha_r^\Gamma$ in Eq.~\eqref{fitfunctionsG} to the theoretical prediction, and performing a two parameters fit in which the quantity $\gamma_r^\Gamma$ appearing in Eq.~\eqref{fitfunctionsG} corresponds to $\Delta\hat F_r^{(1)}$. In Fig.~\ref{fig:DFA} we compare our prediction for $\Delta\hat F_r^{(1)}$, given in Table~\ref{tab:J}, with the results of our fit procedure for $\gamma_r^{\Gamma}$ for $-\frac{1}{2}<r\leq 5$. Observe that the numerical evaluation of the single contribution $\Delta\hat F_r^{(1)}$ is not possible for $r=0$. In this case, the result of our fit for the $O(N^{-1})$ correction was $\beta^\Gamma_r+ \gamma_r^\Gamma=-0.97(4)$, to be compared with the theoretical prediction $N(\Delta\hat F^{(1)}_0\!+\!\Delta\hat F^{F}_0\!+\!\Delta\hat F^{T}_0)=-0.998354732\dots$.

\section{Conclusions}\label{sec:conc}
In the present paper we have discussed the finite-size corrections in the random assignment problem for a generic distribution law $\rho_r(w)$ for the weights in the form of Eq.~\eqref{pr}. We have shown that, for $r> 0$ and $\eta_1\neq 0$, the first finite-size correction scales as $O(N^{-\frac{1}{r+1}})$ and it is proportional to $\eta_1$. In particular, the integrals $J_r^{(0)}$, $J_r^{(r)}$ and $J_r^{(r+3)}$ are positive quantities (see Appendix \ref{app:G}). Therefore, independently of $r$, the corrections $\Delta\hat{F}_r^T$ and $\Delta\hat{F}_r^F$ are negative, while $\Delta\hat{F}_r^{(1)}$ has opposite sign with respect to $\eta_1$, so that, for example, it is positive for the law $\rho_r^\Gamma$, while it vanishes for the law $\rho_r^\text{P}$. We also provided a general expression for the coefficients of the $O(N^{-\frac{1}{r+1}})$ and $O(N^{-1})$ corrections. Moreover, we have shown that, if $\lim_{r\to+\infty}\rho_r=\rho$, then in general
\begin{equation}\label{limiti}
\lim_{r\to\infty}\lim_{N\to\infty}\hat E_r(N) \neq \lim_{N\to\infty}\lim_{r\to\infty}\hat E_r(N).
\end{equation}
We have finally numerically verified our results, by a numerical integration of our formulas and a comparison with simulations.

The $O(N^{-\frac{k}{r+1}})$ corrections appearing in Eq.~\eqref{resume}, for $2\leq k\leq [r]+1$, remain to be computed. As discussed above, in the $r\to\infty$ limit, it is expected that all these finite size corrections contribute to the leading term, justifying the noncommutativity of the limits in Eq.~\eqref{limiti}.

\section{Acknowledgments}
The authors thank Giorgio Parisi for many discussions and suggestions. The authors also thank Carlo Lucibello and Marco Gherardi for discussions on the numerical solution of the integral equations involving $\hat G_r$. M.P.D.~thanks the LCM for the extensive use of their computational resources. The work of G.S.~was supported by the Simons Foundation (Grant No.~454949, Giorgio Parisi).
\appendix
\section{Evaluation of $z[Q]$ on the saddle point and analytic continuation for $n\to 0$}\label{app:zq}
Let us evaluate now the quantity $z[Q]$ on the saddle point. Using the fact that, for any analytic function $f$,
\begin{equation}
 \int_0^{2 \pi} \frac{\dd \lambda} {2 \pi}  \e^{i \lambda} f\left(\e^{-i \lambda} \right)
 = \oint \frac{\dd\xi}{2 \pi i} \frac{f(\xi)}{\xi^2} =  \left. \frac{\dd f}{\dd\xi} \right|_{\xi=0},
\end{equation}
we can write
\begin{multline}
\left[\prod_{a=1}^n\int_0^{2 \pi} \frac{\dd \lambda^a} {2 \pi}  \e^{i \lambda^a} \right]\exp \left\{ \sideset{}{'} \sum_{\alpha\in {\cal P}([n])} q_{|\alpha|}\e^{- i \sum_{b\in \alpha} \lambda^b}   \right\}  \\
= \left. \frac{\partial^n}{\partial\xi_1 \cdots \partial\xi_n} \right|_{\xi_1=\cdots = \xi_n = 0} 
\exp \left\{  \sideset{}{'} \sum_{\alpha\in {\cal P}([n])} q_{|\alpha|} \prod_{b\in \alpha} \xi_b   \right\} \\
=\sum_{\boldsymbol \alpha} \prod_{\alpha_i \in \boldsymbol{\alpha}} q_{|\alpha_i|},
\end{multline}
where $\boldsymbol \alpha = \{ \alpha_i\}_i $ and $\alpha_i\in {\cal P}([n])$ are disjoint subsets whose union is $[n]$; however,
\begin{multline}
\sum_{\boldsymbol \alpha}\prod_{\alpha_i \in \boldsymbol{\alpha}} q_{|\alpha_i|}= 
\sum_{m=1}^n \sum_{\substack{k_1, \dots, k_m \\ k_1+\dots+k_m =n}}
{n \choose k_1 \dots k_m}\frac{q_{k_1} \dots q_{k_m}}{m!} \\
= \left. \left(\frac{\dd}{\dd t}\right)^n \sum_{m=0}^\infty \frac{1}{m!}  \sum_{k_1, \dots, k_m} \frac{q_{k_1}\dots q_{k_m}}{k_1! \dots k_m!} t^{k_1+\dots+k_m}  \right|_{t=0} \\
= \left. \left(\frac{\dd}{\dd t}\right)^n \sum_{m=0}^\infty \frac{1}{m!} \left( \sum_{k=1}^\infty q_k \frac{t^k}{k!} \right)^m \right|_{t=0} \\
= \left. \left(\frac{\dd}{\dd t}\right)^n \exp \left( \sum_{k=1}^\infty q_k \frac{t^k}{k!} \right) \right|_{t=0}.
\end{multline}
To perform the analytic prolongation, we prove now that, if $f(0)=1$, then
\begin{equation}
\lim_{n \to 0}\frac{1}{n} \ln\left[\left. \left(\frac{d}{dt}\right)^n f(t) \right|_{t=0} \right] = \int_{-\infty}^{+\infty}\left[ \e^{-\e^l} - f(- \e^l)\right]\dd l.\label{A}
\end{equation}
This fact can be seen observing that, for $n\to 0$,
\begin{multline}
\left(\frac{\dd}{\dd t}\right)^n f(t) =\left. f \left( \frac{\partial}{\partial J} \right)J^n \e^{J t} \right|_{J=0} \\
\approx \left. f(t) + n \,  f \left( \frac{\partial}{\partial J} \right) \, \ln J \,  \e^{J t} \right|_{J=0} \\
=\left. f(t) + n \,  f \left( \frac{\partial}{\partial J} \right) \, \int_0^\infty \frac{\dd s}{s} \, \left( \e^{-s} - \e^{-s J} \right) \, \e^{J t} \right|_{J=0} \\
=f(t) + n  \int_0^\infty \frac{\dd s}{s} \,\left[ \e^{-s}\, f(t)  - f(t-s) \right] \, .
\end{multline}
By the change of variable $s=\e^l$, Eq.~\eqref{A} follows.
\section{Asymptotic behaviour of the function $B_r$}\label{app:B}
In this appendix we study the asymptotic behavior for large $\lambda$ of the function $B_r(\lambda x)$. By definition in Eq.~\eqref{Br}
\begin{equation}
\frac{1}{\lambda^r} B_r(\lambda x)\coloneqq \sum_{k=1}^\infty (-1)^{k-1}\frac{\e^{\lambda xk}}{(\lambda  k)^r(k!)^2}, 
\end{equation}
so that
\begin{multline}
\frac{B_r(\lambda x)}{\lambda^r} 
=- \frac{1}{\Gamma(r)}\int_0^\infty t^{r-1} \sum_ {k=1}^\infty  \frac{(-1)^k }{(k!)^2}e ^{\lambda (x-t)k} \dd t\\
=- \frac{1}{\Gamma(r)} \int_0^\infty t^{r-1} \left\{ J_0\left[ 2 \e^{\frac{1}{2}\lambda(x-t)}\right] -1 \right\}\dd t\\
\xrightarrow{\lambda\to+\infty}
\frac{1}{\Gamma(r)} \int_0^\infty t^{r-1}\theta( x-t )\dd t =\frac{x^r}{\Gamma(r+1)}\theta(x),
\end{multline}
where we have used the fact that
\begin{equation}
J_0(x)\coloneqq\sum_{m=0}^\infty \frac{(-1)^m }{(m!)^2} \left( \frac{x}{2} \right)^{2m} =
\begin{cases}
1 & {\rm when \; } x \to  0 , \\
0             & {\rm when \; } x \to + \infty
\end{cases}
\end{equation}
is the Bessel function of zeroth order of the first kind.
\section{Some properties of the function $\hat G_r$}\label{app:G}
In this appendix we give some properties of the function $\hat G_r$, defined by the integral equation \eqref{Gr}. From the definition, we have that, for $0\leq \alpha<\beta+1$ and $r>-1$,
\begin{multline}\label{Gderivata}
\hat{G}^{(\alpha)}_r(l)\coloneqq\fder{\alpha}{l}\hat{G}_r(l) =\\
= \int_{-\infty}^{+\infty} \frac{(l+y)^{r-\alpha}}{\Gamma(r-\alpha+1)}\e^{-\hat{G}_r(y)}\theta(l+y)\dd y.
\end{multline}
Observe that
\begin{equation}
\hat{G}^{(\alpha)}_r(l)\geq 0\quad \text{for}\quad 0\leq\alpha< r+1.
\end{equation}
In this equation we have used the fact that, for $0\leq\alpha<\beta+1$, we have \cite{Podlubny1999}
\begin{equation}
\fder{\alpha}{t}\left[\frac{t^\beta}{\Gamma(\beta+1)}\theta(t)\right]=\frac{t^{\beta-\alpha}}{\Gamma(\beta-\alpha+1)}\theta(t).
\end{equation}
In particular, for $\alpha=r$ we have the simple relation
\begin{equation}\label{intexpg}
\hat{G}^{(r)}_r(l)\coloneqq\fder{r}{l}\hat{G}_r(l) = \int_{-\infty}^\infty \e^{-\hat{G}_r(y)}\theta(y+l)  \dd y.
\end{equation}
Moreover, for $0\leq \alpha<r+1$,
\begin{equation}
\lim_{l \to - \infty} \hat{G}^{(\alpha)}_r(l)=0.
\end{equation}
From Eq.~\eqref{intexpg}
\begin{equation}\label{Grderr}
\hat{G}^{(r+1)}_r(l) = \e^{-\hat{G}_r(-l)}\geq 0\Rightarrow \lim_{l \to + \infty} \hat{G}^{(r+1)}_r(l) = 1.
\end{equation}
The relations above imply that
\begin{equation}
J^{(\alpha)}_r\coloneqq\int_{-\infty}^{+\infty}\hat G_r(-u)\fder{\alpha}{u}\hat G_r(u)\dd u>0,\quad 0\leq \alpha<r+1.
\end{equation}
Similarly, for $0<k<r+1$ an integer,
\begin{multline}
J^{(r+k+1)}_r\coloneqq\int_{-\infty}^{+\infty}\hat G_r(-u)\fder{r+k+1}{u}\hat G_r(u)\dd u\\
=\int_{-\infty}^{+\infty}\hat G_r(-u)\frac{\dd^k}{\dd u^k}\e^{-\hat G_r(-u)}\dd u\\
=\int_{-\infty}^{+\infty}\hat G_r^{(k)}(u)\e^{-\hat G_r(u)}\dd u\geq 0.
\end{multline}
For large $l$ we have
\begin{align}
\hat{G}_r(l) \approx  & \frac{l^{r+1}}{\Gamma(r+2)},\\
\hat{G}_r(-l) \approx  &\exp\left[- \frac{l^{r+1}}{\Gamma(r+2)} \right].
\end{align}
As anticipated, an exact solution is available in the $r=0$ case. In particular, for $r=0$, the second derivative
\begin{equation}
\hat{G}^{(2)}_0(l) = \e^{-\hat{G}_0(-l)}\hat{G}^{(1)}_0(-l) =  \hat{G}^{(1)}_0(l)\hat{G}^{(1)}_0(-l)
\end{equation}
is an even function of $l$,
\begin{equation}
\hat{G}^{(2)}_0(l) - \hat{G}^{(2)}_0(-l) = 0\Rightarrow\hat{G}^{(1)}_0(l) + \hat{G}^{(1)}_0(-l) =  c,
\end{equation}
with the constant $c=1$ by evaluating the left-hand side in the limit of infinite $l$ and
\begin{equation}
\hat{G}^{(1)}_0(0) = \e^{-\hat{G}_0(0)} = \frac{1}{2}. \label{ic}
\end{equation}
Then we have that
\begin{equation}
\hat{G}_0(l) - \hat{G}_0(-l) =  l\Rightarrow\hat{G}^{(1)}_0(l) = \e^{-\hat{G}_0(-l)} = \e^{\, l-\hat{G}_0(l)},
\end{equation}
which means that
\begin{equation}
\frac{\dd}{\dd l} \e^{\hat{G}_0(l)} = \e^l\Rightarrow \e^{\hat{G}_0(x)} - \e^{\hat{G}_0(0)}   = \e^x - 1,
\end{equation}
where we have used the initial condition at $l=0$, that is, because of~Eq.~\eqref{ic},
\begin{equation}
\e^{\hat{G}_0(x)} = 1+ \e^x\Rightarrow \hat{G}_0(x)=\ln(1+ \e^x).\label{Gesatta}
\end{equation}
\section{Evaluation of the integrals in the $r=0$ case} \label{app:J}
To explicitly evaluate some of the integrals above, let us introduce the polygamma function
\begin{equation}
\psi_m(z)\coloneqq \frac{\dd^{m+1} }{\dd z^{m+1}} \ln \Gamma(z) = (-1)^{m+1}\int_0^\infty\frac{t^m \e^{-zt}}{1 - \e^{-t}}\dd t,
\end{equation}
which satisfies the recursion relation
\begin{equation}
\psi_m(z+1) = \psi_m(z) + (-1)^m \frac{m!}{z^{m+1}},
\end{equation}
which, for a positive integer argument and assuming $m\geq 1$, leads to
\begin{equation}
\frac{\psi_m(k)}{(-1)^{m+1}m!} = \zeta(m+1) -\sum_{r=1}^{k-1}\frac{1}{r^{m+1}} =  \sum_{r=k}^\infty \frac{1}{r^{m+1}}.
\end{equation}
For $m=0$ this implies
\begin{equation}
\psi_0(k) =   -\gamma_E + H_{n-1}\Rightarrow \psi_0(1) = - \gamma_E,
\end{equation}
with $\gamma_E$ is Euler's gamma constant and
\begin{equation}
H_{n}\coloneqq \sum_{k=1}^n \frac{1}{k}
\end{equation}
are the {harmonic numbers}.

With these considerations in mind and using Eq.~\eqref{Gesatta}, we have that
\begin{multline}
J_0^{(1)} \coloneqq \int_{-\infty}^{+\infty} \frac{\ln (1 + \e^y)}{1 + \e^y}\dd y\\
= \int_{0}^{+\infty}\frac{t\, \e^{-t}}{1- \e^{-t}} \dd t= \psi_1(1) = \zeta(2) = \sum_{k \ge 1} \frac{1}{k^2}  = \frac{\pi^2}{6}.
\end{multline}
Then we compute
\begin{multline}
J_0^{(0)}\coloneqq \int_{-\infty}^{+\infty}\dd y\,\frac{1}{1 + \e^y}\int_{-\infty}^y \dd u\, \ln (1 + \e^u)\\
=\int_{0}^{+\infty} \dd t\, \frac{\e^{-t}}{1-\e^{-t}}\int_{0}^t \dd w\frac{w}{1-\e^{-w}}
=-\int_{0}^{+\infty} dt\, t\, \frac{\ln(1-\e^{-t})}{1-\e^{-t}}  \label{167} \\
=\sum_{k=1}^\infty\frac{1}{k}\int_{0}^{+\infty} \frac{ t \e^{-k t}}{1-\e^{-t}}\dd t=\sum_{k=1}^\infty\frac{1}{k}\psi_1(k).
\end{multline}
We remark now that
\begin{multline}
\sum_{k\ge 1} \frac{\psi_1(k)}{k}=\sum_{k=1}^\infty\sum_{r=0}^\infty\frac{1}{k}\frac{1}{(r+k)^2}\\
=   \, \sum_{s \ge 1} \,   \sum_{k = 1}^s \, \frac{1}{k}\, \frac{1}{s^2} 
=   \, \sum_{s \ge 1} \,  \frac{1}{s^2}   H_s.
 \end{multline}
Applying now the identity
\begin{equation}
\sum_{s=1}^\infty\frac{H_s}{s^2}= 2\zeta(3),
\end{equation}
discovered by Euler, we recover the result obtained by \textcite{Ratieville2002}
%
\begin{equation}
J_0^{(0)} = 2\zeta(3) =- \psi_2(1).
\end{equation}

To finally evaluate $J^{(3)}_0$, we remark now that
\begin{equation}
\int_{-\infty}^{+\infty}  \frac{\dd y}{1 + \e^y}\frac{\dd}{\dd y}  \ln (1 + \e^y) =
-\int_{-\infty}^{+\infty} \dd y \frac{\dd}{\dd y}  \frac{1}{1 + \e^y} = 1.
\end{equation}
Then, as
\begin{equation}
\frac{\dd^2}{\dd y^2}\ln (1 + \e^y)  = \frac{\dd}{\dd y}  \ln (1 + \e^y)  - \left[ \frac{\dd}{\dd y}  \ln (1 + \e^y) \right]^2,
\end{equation}
we have
\begin{multline}
J^{(3)}_0=\int_{-\infty}^{+\infty} \dd y \, \frac{1}{1 + \e^y} \frac{\dd^2}{\dd y^2}\ln (1 + \e^y) =\\
=- \int_{-\infty}^{+\infty} \dd y \, \left(\frac{\dd}{\dd y} \frac{1}{1 + \e^y}\right)\frac{\dd}{\dd y}  \ln (1 + \e^y)\\
= \int_{-\infty}^{+\infty} \dd y \, \frac{1}{1 + \e^y}  \left[ \frac{\dd}{\dd y}  \ln (1 + \e^y) \right]^2=\frac{1}{2}.\label{12}
\end{multline}

\begin{widetext}
\section{Calculation of $\Delta F_r^T$}\label{app:FT}
To evaluate explicitly $\Delta F_r^T$, let us start from Eq.~\eqref{DFT}, \begin{multline}
\Delta \hat{F}_r^T = \frac{1}{2 \hat{\beta} N} \sum_{s=1}^\infty \sum_{t=1}^\infty (-1)^{s+t -1} \frac{(s+t-1)!}{s!t!} \frac{\hat{g}_s \hat{g}_t}{\hat{g}^2_{s+t} } q_{s+t}^2 
= \frac{\eta_0\Gamma(r+1)}{2 \hat{\beta}^{r+2} N} \sum_{s=1}^\infty \sum_{t=1}^\infty \frac{(-1)^{s+t -1}}{s! t!} \left( \frac{t+s}{s\, t} \right)^{r+1}\int_{-\infty}^{+\infty} \e^{-G_r(y)} \e^{y (s+t)} q_{s+t}\dd y\\
= \frac{\eta_0\Gamma(r+1)}{2 \hat{\beta}^{r+2} N} \sum_{k=2}^\infty \sum_{s =1}^{k-1} \frac{(-1)^{k-1}}{s! \,(k-s)!}  \frac{k^{r+1}}{s^{r+1}(k-s)^{r+1}}\int_{-\infty}^{+\infty}\e^{-G_r(y)} \e^{yk} q_{k} \dd y, \end{multline}
and, in order to perform the sum over $s$, we introduce integral representations
\begin{equation}
\sum_{s=1}^{k-1} \frac{1}{s!(k-s)!}\frac{1}{s^{r+1}(k-s)^{r+1}}
=\sum_{s =1}^{k-1} \binom{k}{s} \int_0^{+\infty} \dd u \int_0^{+\infty} \dd v\, \frac{u^r v^r \e^{-s u} \e^ {-(k-s)v}}{k! \Gamma^2(r+1)}
=\int_0^{+\infty} \dd u \int_0^{+\infty} \dd v\, u^r v^r \frac{\left(\e^{-u}\! +\! \e^{-v}\right)^k\! -\! \e^{- u k}\! - \e^{- v k}}{k!\Gamma^2(r+1)}.
\end{equation}
Observing now that the value $k=1$ can be included in the sum over $k$ and defining
\begin{equation}
h\coloneqq \frac{\hat \beta}{\left[\eta_0 \Gamma(r+1) \right]^{1/(r+1)}},
\end{equation}
we can write
\begin{multline}
\frac{2 \hat{\beta}^{r+2}\Gamma(r+1)N}{\eta_0}\Delta \hat{F}_r^T
=\sum_{k=1}^\infty\frac{(-1)^{k-1}}{k!} q_k  \int_{-\infty}^{+\infty}\dd y\, \e^{-G_r(y)}\fder{r+1}{y}\e^{yk} \int_0^{+\infty} \dd u \int_0^{+\infty}\dd v\, u^r v^r \left[\left(\e^{-u} + \e^{-v}\right)^k - \e^{- u k} - \e^{- v k}\right]\\
= \int_{-\infty}^{+\infty} \dd y \,\e^{-G_r(y)}\fder{r+1}{y}\left\{\int_0^{+\infty} du \int_0^{+\infty} dv\, u^r v^r  \left[ G_r\left( y + \ln \left( \e^{-u} + \e^{-v}\right) \right) - G_r(y-u) -  G_r(y-v)\right]\right\}\\
=2h^r\int_{-\infty}^{+\infty} \dd y \, \e^{-\hat{G}_r(y)}\fder{r+1}{y}\left\{\int_0^{+\infty} \dd u \int_0^{u} \dd v\, u^r v^r  \left[ G_r\left( h(y-v) + \ln \left(\e^{-h(u-v)} + 1\right) \right) -\hat G_r(y-u) -\hat G_r(y-v)\right]\right\}.\end{multline}
This implies that, for $h\to\infty$,
\begin{multline}
\Delta \hat{F}_r^T= \frac{1}{N\eta_0^{\frac{1}{r+1}}\Gamma^{2+\frac{1}{r+1}}(r+1)}\int_{-\infty}^{+\infty} \dd y \, \e^{-\hat{G}_r(y)}\fder{r+1}{y}\left[\int_0^{+\infty} \dd u \int_0^{u} \dd v\, u^r v^r  \hat G_r(y-u)\right]\\
=-\frac{1}{N\eta_0^{\frac{1}{r+1}}\Gamma^{2+\frac{1}{r+1}}(r+1)}\frac{1}{r+1}\int_{-\infty}^{+\infty} \dd y \, \e^{-\hat{G}_r(y)}\fder{r+1}{y}\left[\int_{-\infty}^y (y-u)^{2r+1}\hat G_r(u)\dd u\right]\\
=-\frac{\Gamma(2r+2)}{N\eta_0^{\frac{1}{r+1}}\Gamma^{3+\frac{1}{r+1}}(r+1)}\frac{1}{r+1}\int_{-\infty}^{+\infty} \dd y \, \e^{-\hat{G}_r(y)}\int_{-\infty}^y (y-u)^{r}\hat G_r(u)\dd u
=-\frac{\Gamma(2r+2)}{N\eta_0^{\frac{1}{r+1}}\Gamma^{2+\frac{1}{r+1}}(r+1)}\frac{1}{r+1}\int_{-\infty}^{+\infty}\hat G_r(-u)\hat G_r(u)\dd u,
\end{multline}
which is exactly Eq.~\eqref{DFTfin}.\end{widetext}
\bibliography{biblio.bib}
\end{document}